\documentclass[twocolumn,showpacs,preprintnumbers,amsmath,amssymb,floatfix]{revtex4}
\usepackage{graphicx}
\usepackage{dcolumn}
\usepackage{bm}
\usepackage{multirow}
\usepackage[dvips]{hyperref}
\usepackage{amsmath}

\begin{document}

\title{Microscopic calculation and local approximation of the spatial dependence 
of the pairing field with bare and induced interactions}
\author{ A.Pastore$^{a,b}$, F.Barranco$^c$, R.A.Broglia$^{a,b,d}$  and E.Vigezzi$^b$}
\affiliation{
$^a$ Dipartimento di Fisica, Universit\`a degli Studi di Milano,via Celoria 16, 20133 Milano, Italy.\\
$^b$ INFN, Sezione di Milano, via Celoria 16, 20133 Milano, Italy.\\
$^c$ Departamento de Fisica Aplicada III, Escuela Superior de Ingenieros, Camino de los Descubrimientos s/n,  
  41092 Sevilla, Spain.\\
$^d$ The Niels Bohr Institute, University of Copenhagen, Blegdamsvej 17, 2100 Copenhagen \O, Denmark.}

\date{\today} 
\begin{abstract}
The bare nucleon-nucleon interaction is essential for the production of pair correlations in nuclei, 
but an important contribution also arises from the induced interaction 
resulting from the exchange of collective vibrations between 
nucleons moving in time reversal states close to the Fermi energy.  
The pairing field resulting from the summed interaction is strongly 
peaked at the nuclear surface.
It is possible to reproduce the detailed spatial 
dependence of this field using either a local approximation which takes fully into account
finite size effects, or  a 
 contact interaction, with parameters which are quite different from those commonly used
in more phenomenological approaches.  
\end{abstract}

\maketitle

\section{Introduction}

Pairing correlations influence in an essential way basic properties of atomic nuclei \cite{Brink&Broglia}.
A consistent approach to describe these correlations employs a bare nucleon-nucleon interaction whose parameters are fitted so as to reproduce the experimental phase shifts (like the $v_{14}$ Argonne potential), and includes medium polarization effects.
The exchange of vibrations between nucleons moving in time reversal states lying close to the Fermi energy has been shown to account in both stable and halo nuclei for a consistent fraction of the pairing gap and of the two-nucleon separation energy \cite{Barranco_EPJA04}-\cite{PRC04_Gori}.
 The coupling of nucleons and vibrations renormalizes in an important way the single-particle properties of atomic nuclei, 
leading to changes in the level densities at the Fermi energy and to a breaking of single particle strength
(dynamical shell model \cite{Mahaux}). 
As a rule, the dynamical shell model phenomena are simply parametrized in terms of an effective mass 
and of spectroscopic factors. In this paper we follow the rule and concentrate  on the detailed study of the spatial dependence 
of the pairing field, without pretending to achieve a precise estimate of the value of the pairing gap and of the 
condensation energy. We plan  to come back to the issue in a future publication, taking the variety of  medium polarization effects into account  in a self-consistent  and unified way  within the framework  of the approach 
employed in \cite{Barranco_EPJA04}, based on the solution of the Nambu-Gorkov equations 
and using renormalized QRPA phonons to describe the collective modes.

The main subject of the present work is the spatial dependence of the pairing field and of the pairing density associated with the neutrons of
$^{120}$Sn associated with the bare and induced pairing interaction.

In atomic nuclei,
the coherence length is a few times larger than  the nuclear radius. Consequently
a simple Local Density Approximation, based on the results obtained in uniform
matter, is not expected to lead to accurate results in the case of the finite system. 
The fact that the  
wavefunction  of the Cooper pair is largely independent of the nuclear interaction, being 
dominated by the spatial dependence of a few orbitals lying  around  the Fermi surface,
testifies to this expectation.
We shall instead parametrize our results
in terms of a local approximation which reproduces the spatial dependence of the pairing field 
resulting from 
the microscopic
calculations. In this way, the presence of the nuclear surface is taken into account  in an effective way.
This will allow us to make a detailed comparison with effective forces commonly
used to calculate pairing correlations, like the Gogny force and zero-range, density dependent interactions.


\section{Solution of HFB equations and the spatial dependence of the pairing field }

We start by performing a Hartree-Fock calculation with the two-body interaction SLy4~\cite{Chabanat} (associated with a $k$-mass $m_k \simeq 0.7m$ at saturation density), obtaining
a set of single-particle energy levels $e_{nlj}$.
Using different pairing interactions, that will be discussed below, we 
then solve in the calculated HF basis  the Hartree-Fock-Bogoliubov (HFB)  equations 
in the pairing channel, 
\begin{equation}\label{HFB}\begin{split}
  (e_{nlj}-e_{F})U^q_{nlj}+\sum_{n'}\Delta_{nn'lj}V^q_{n'lj}=E^{q}_{lj}U^q_{nlj}, \\
\sum_{n'}\Delta_{nn'lj}U^q_{n'lj} -(e_{nlj}-e_{F})V^q_{nlj}  =E^{q}_{lj}V^q_{nlj},
\end{split}
\end{equation}
where $E^q_{lj}$ denotes the quasiparticle energy, $U^q_{nlj}$ and $V^q_{nlj}$
are the associated amplitudes, while $e_{nlj}$  denote the HF single-particle energies 
and $e_F$ is the Fermi energy.
\noindent 
The calculation are performed in a spherical box of radius $R_{box}$ = 15 fm.
From the quasiparticle amplitudes one can construct the abnormal density, also 
referred to as the Cooper pair wavefunction:
\begin{eqnarray}\label{a:density:r} 
\Phi(\vec{r}_1,\vec{r}_2)=\sum_{qnn'lj} \frac{2j+1}{2} 
U^q_{nlj}V^q_{n'lj}\psi_{nn'lj}(\vec{r}_1,\vec{r}_2),
\end{eqnarray}
where $\psi_{nn'lj}(\vec{r}_1,\vec{r}_2)=[\phi_{nlj}(\vec{r}_1)\phi_{n'lj}(\vec{r}_2)]_{00}$ is the wavefunction of two neutrons
coupled to $J=0$.  
We shall only consider the $S=0$ component of $\Phi$,\; $\Phi^{S=0}$, which is by far the dominant one.
We then insert in Eq.(\ref{a:density:r}), in place of $\psi_{nn'lj}$, the function 
\begin{eqnarray}\label{2:wave}
\psi_{nn'lj}^{S=0}(\vec{r}_1,\vec{r}_2) = \langle\vec{r}_1,\vec{r}_2|nn'lj;J=0\rangle_{S=0} \nonumber \\
= \frac{1}{4\pi}\phi_{nlj}(r_1)\phi_{n'lj}(r_2)P_l(\cos\theta_{12}),
\end{eqnarray}
where $P_l$ is a Legendre polynomial.

The matrix elements of the pairing field
are obtained self-consistently from the abnormal density using the state dependent gap 
\begin{equation}
 \Delta_{nn'lj}= - \langle nn'lj;J=0|v|\Phi \rangle,
\label{deltann}
\end{equation}
where $v$ is the pairing interaction. 

\vspace{5mm}
In the present paper we shall determine the 
spatial dependence of the pairing gap, using a simplified version of 
the formalism adopted in ref.~\cite{Barranco_EPJA04}, which is convenient 
to make contact with phenomenological approaches (cf. the discussion in the 
Appendix of ref. \cite{PRC05_Barranco}). The total interaction is given
by the sum of the bare nucleon-nucleon interaction, here taken to be 
the Argonne $v_{14}$ interaction $v_{Arg}$, and of the interaction induced by the 
exchange of vibrations $v_{ind}$.
We shall renormalize the matrix elements $v_{Arg} + v_{ind}$ of the total interaction, using matrix elements $v_{Arg + ind}$ which take into account fragmentation and self energy effects:   
\begin{eqnarray}\label{Int:arg+indotta}
&&\langle \nu'_1 m' \nu'_2  \bar{m'}|v_{Arg + ind}| \nu_1 m  \nu_2 \bar{m}  \rangle = \nonumber \\
&& Z  \langle \nu'_1 m' \nu'_2  \bar{m'}|v_{Arg} + v_{ind}| \nu_1 m  \nu_2 \bar{m}  \rangle
\end{eqnarray}
\noindent where $\nu$ stands for $\{nlj\}$, $|\bar{m}>$ denotes the time reversed state, 
$|\bar{m}> = (-1)^{m+j}|-m>$, and   $Z$ denotes an average value of the quasiparticle strength at the Fermi energy.
In the following we shall use the typical value $Z = 0.7$~\cite{Terasaki_02,Mahaux}. 
Vertex corrections are not considered, 
because their contribution to the pairing gap has been found to be very small in the detailed calculation performed solving 
the Nambu-Gorkov equation \cite{Barranco_EPJA04}. 

\begin{figure*}
\begin{center}
\includegraphics[width=7.5cm,angle=-90]{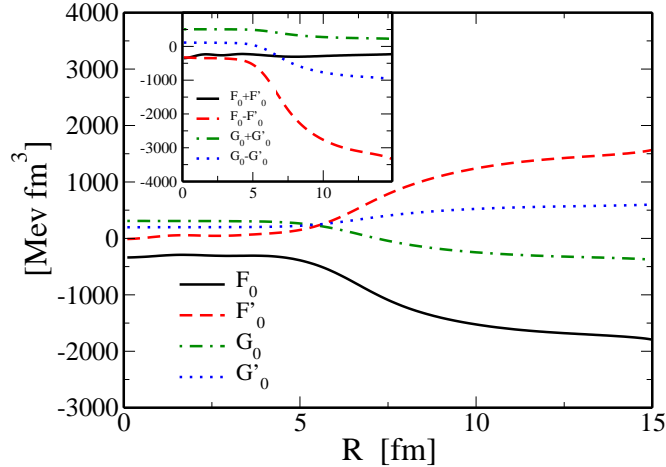}
\end{center}
\caption{ Landau-Migdal parameters associated with the SLy4 force, calculated as a function of the distance from 
the center of the nucleus in $^{120}$Sn.}
\label{Landau}
\end{figure*}

\begin{figure*}
\begin{center}
\includegraphics[width=7.5cm]{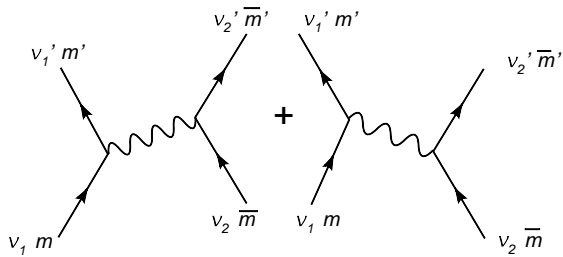}
\end{center}
\caption{Diagrams showing the exchange of a vibration between two pairs of levels coupled to $J=0$.}
\label{diagram}
\end{figure*}

\begin{figure}[htb]
\begin{center}
\includegraphics*[width=7.5cm,angle=-90]{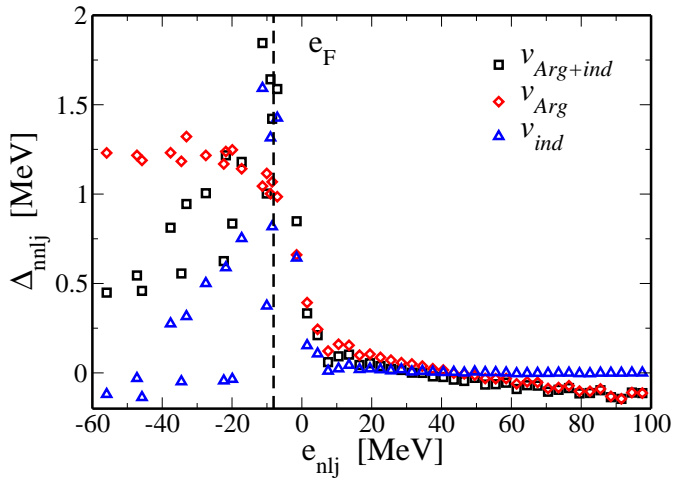}
\end{center}
\caption{\footnotesize{Diagonal matrix elements $\Delta_{nnlj}$ as a function of the single 
particle energy $e_{nlj}$.
The squares, diamonds and triangles refer to  the gaps obtained respectively with the Argonne plus induced 
interaction $v_{Arg + ind}$, with the Argonne interaction $v_{Arg}$ and with the induced interaction $v_{ind}$. 
The vertical dashed line indicates the position of the Fermi energy, that turns out to be 
almost the same for the three calculations.}}
\label{Delta:matrix:el}
\end{figure}

\begin{figure*}[!t]
\begin{center}
\includegraphics*[width=6.3cm,angle=-90]{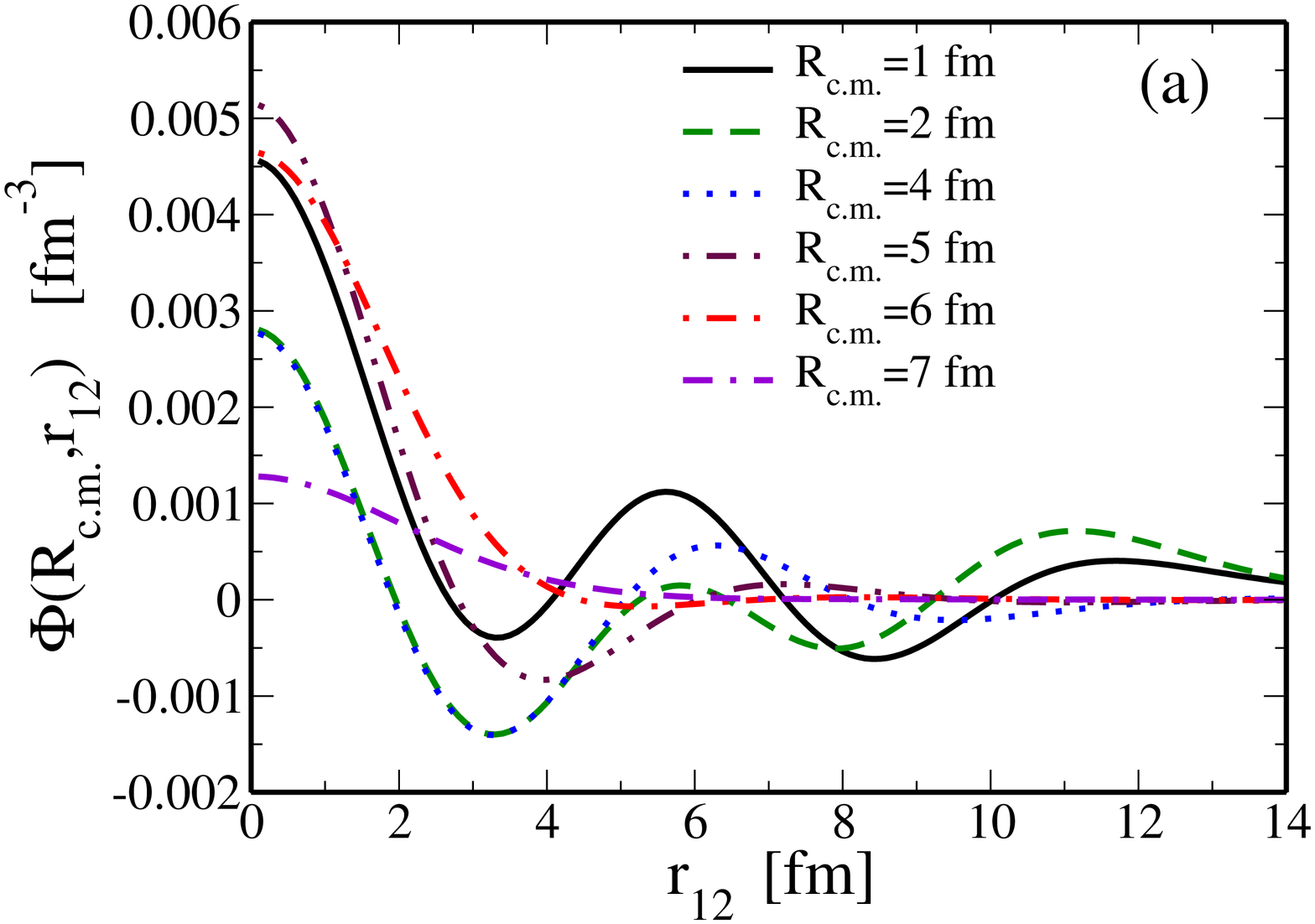}
\hspace{0.075\textwidth}
\includegraphics*[width=6.3cm,angle=-90]{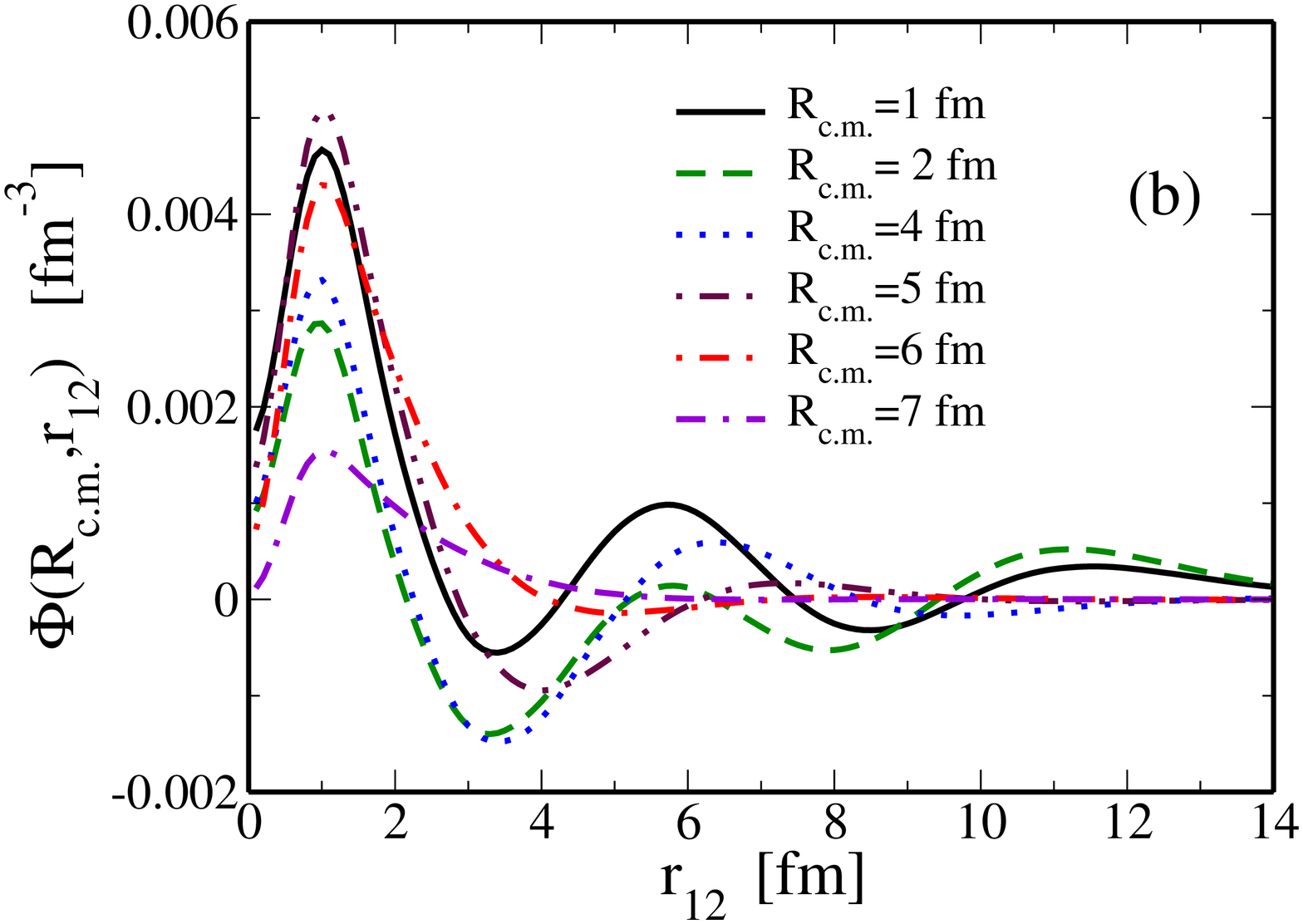}
\end{center}
\caption{\footnotesize{Abnormal density $\Phi(R_{c.m.},r_{12})$ 
for fixed values of $R_{c.m.}$. 
In (a) we show the results calculated with the induced pairing interaction, 
in (b) those obtained with Argonne pairing interaction.}}
\label{Phi:Rr}
\end{figure*}

\begin{figure*}[!t]
\begin{center}
\includegraphics*[width=6.3cm,angle=-90]{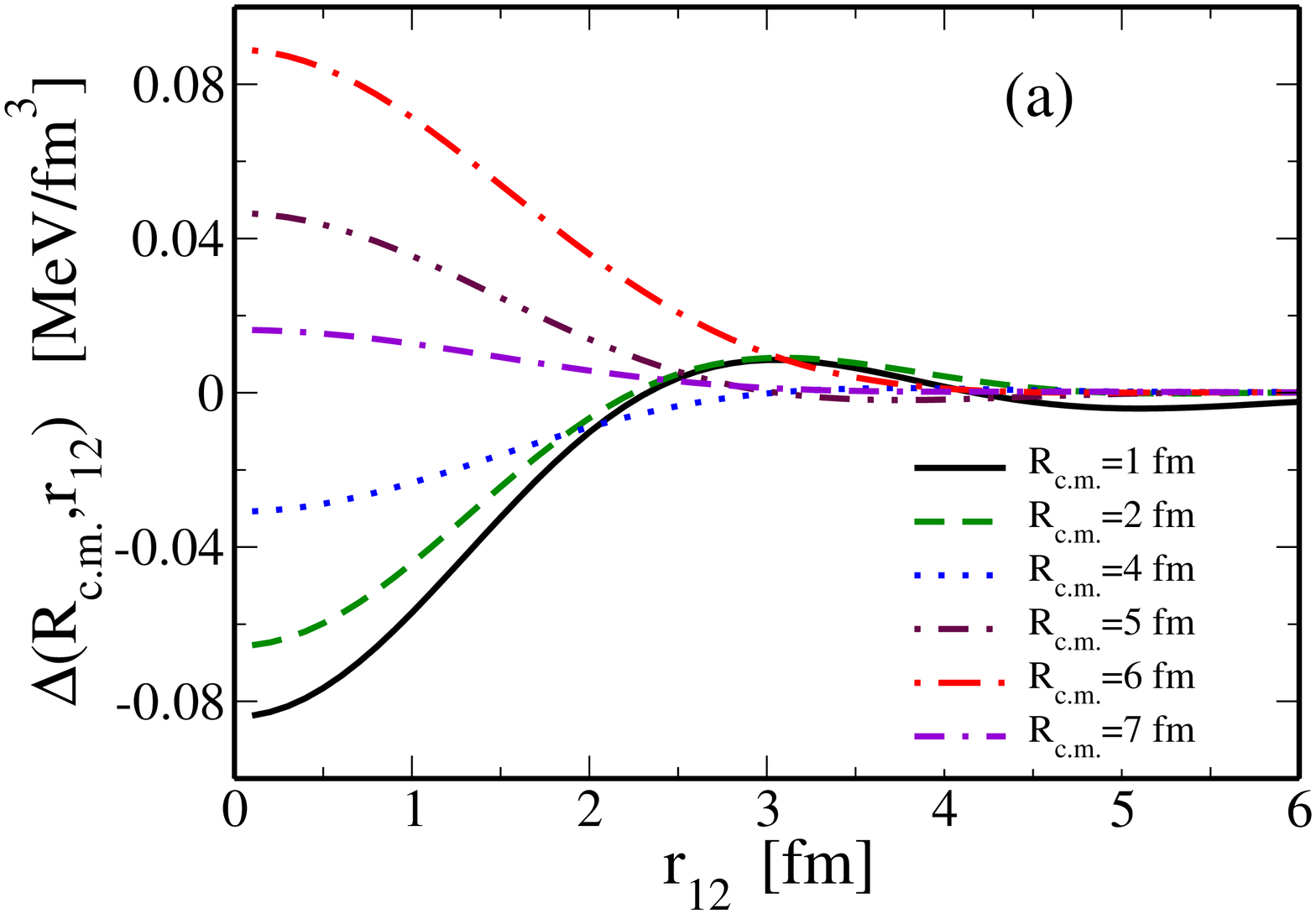}
\hspace{0.075\textwidth}
\includegraphics*[width=6.3cm,angle=-90]{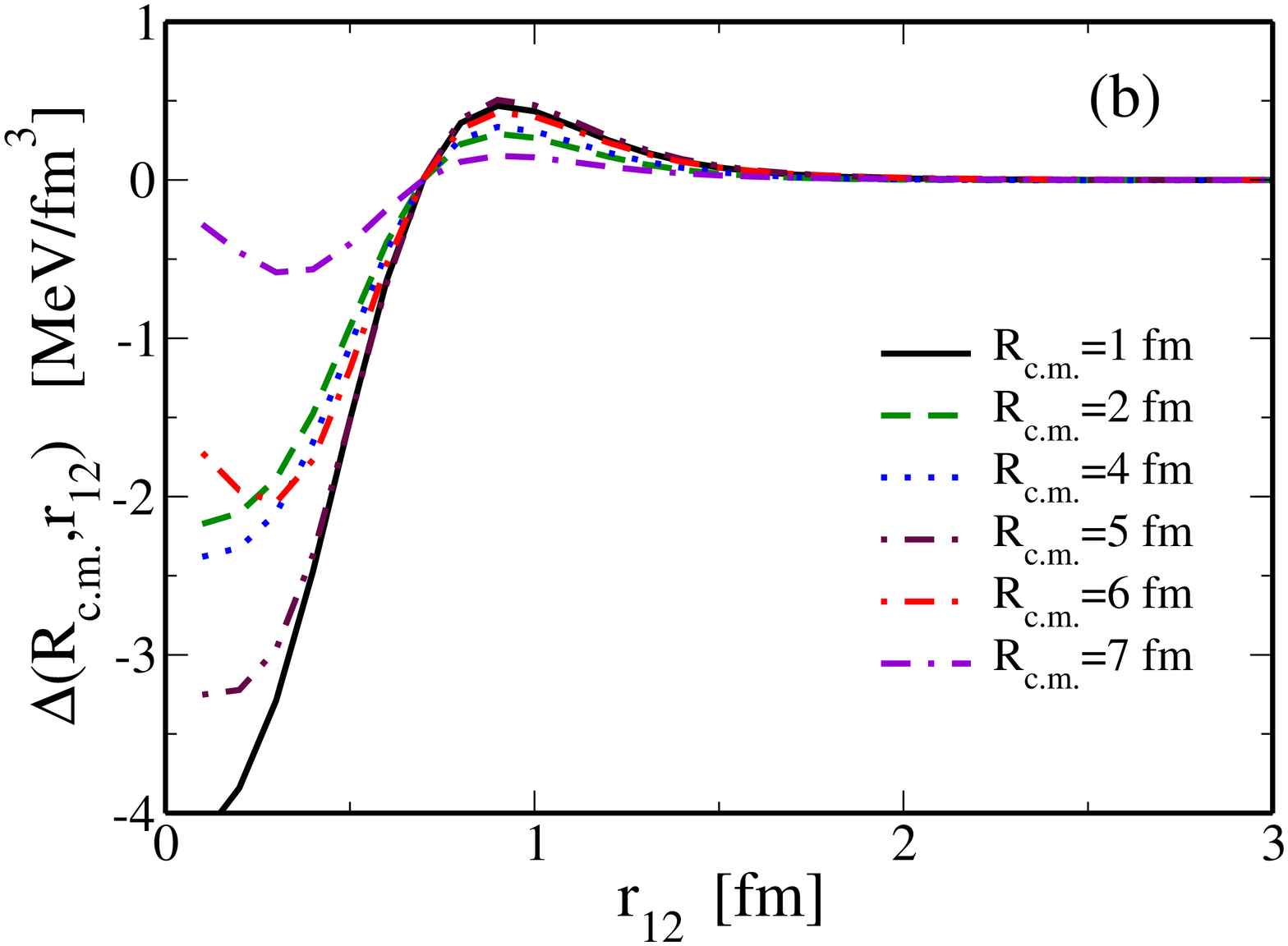}
\end{center}
\caption{\footnotesize{Pairing field $\Delta(R_{c.m.},r_{12})$ 
for fixed values of $R_{c.m.}$.
In the left panel we show the results calculated with the induced pairing interaction, 
in the right panel those obtained with the Argonne pairing interaction.}}
\label{Delta:Rr}
\end{figure*}

The matrix elements of the interaction induced by the exchange of a vibration will be calculated evaluating
the diagrams shown in Fig. \ref{diagram}, 
using the same formalism already employed in ref. \cite{Gori_Sn120}: 
\begin{eqnarray}\label{VIND}
&& \langle \nu'_1 m' \nu'_2 \bar{m'}|v_{ind}| {\nu_1 m}{\nu_2 \bar{m}} \rangle  = \nonumber \\
&& \sum_{J^{\pi}Mi} \frac{(f+g)_{\nu_1 m;J^{\pi}Mi}^{\nu'_1 m'} (f-g)_{\nu_2 m;J^{\pi}Mi}^{\nu'_2 m'}}
  {E_0 - (|e_{\nu'_1}-e_F| + |e_{\nu_2}-e_F| + \hbar\omega_{J^{\pi}i})} \nonumber. \\ 
&& \sum_{J^{\pi}Mi} \frac{(f+g)_{\nu_1 m;J^{\pi}Mi}^{\nu'_1 m'} (f-g)_{\nu_2 m;J^{\pi}Mi}^{\nu'_2 m'}}
  {E_0 - (|e_{\nu_1}-e_F| + |e_{\nu'_2}-e_F| + \hbar\omega_{J^{\pi}i})}+ \nonumber \\
 \end{eqnarray}
\noindent 
The index $i$ labels the exchanged vibrational modes, having a given  angular momentum  and parity $JM^{\pi}$,  and an energy $\hbar \omega_{J^{\pi}i}$. The modes 
have been  calculated in the Quasiparticle Random Phase Approximation 
(QRPA), using the same SLy4 interaction already employed to calculate the mean-field, with the exception
of the spin-orbit and of  the Coulomb part \cite{Colo}.
$E_0$ is the pairing correlation energy of a  Cooper pair, a quantity  
which is of the order of $-2\Delta_F$, where $\Delta_F$ is the average value of the gap close to the Fermi energy. 
In Eq. (\ref{VIND}) $f$ and $g$ denote the particle-vibration coupling vertices associated
with the spin-independent and spin-dependent parts of the residual interaction respectively,
\begin{eqnarray}
&& v_{ph}(\vec{r},\vec{r}')= \delta(\vec{r}-\vec{r}') \times \nonumber\\
&& \times \left\{
\left[F_0+F'_0\vec{\tau}\cdot\vec{\tau}'\right] 
+\left[\left(G_0+G'_0\vec{\tau}\cdot\vec{\tau}'\right)
\vec{\sigma}\cdot\vec{\sigma}' \right]
\right\}.
\label{residual}
\end{eqnarray}

In the calculation of the particle-vibration coupling we neglected the momentum-dependent part of the interaction (this part is instead  taken into account
in the QRPA calculation). 
The vertex $f$ is given by 
\begin{eqnarray}
&& f_{\nu m ; J^{\pi}Mi}^{\nu' m' } = i^{l-l'}
\langle j'm'| (i)^JY_{JM}|jm\rangle  \times \nonumber \\
&& \int dr \varphi_{\nu'}[(F_0+F_0')\delta\rho_{J^{\pi}n}^{i}
+(F_0-F_0')\delta\rho_{J^{\pi}p}^{i}] \varphi_{\nu},
\label{eqn:3}
\end{eqnarray}
where $F_0,F_0'$ are the generalized Landau-Migdal parameters associated
with the SLy4 force and 
controlling the isoscalar and isovector spin-independent channels,
while $\delta\rho^i_{J^{\pi}n}$ and $\delta\rho^i_{J^{\pi}p}$ are the neutron and proton
contributions to the transition densities
given by 
\begin{equation}
\begin{array}{c}
\delta \rho_{J^{\pi}}^{i}(r)=\frac{1}{\sqrt{2J+1}}\sum_{\nu_1,\nu_2}
(X_{\nu_1,\nu_2}(i,J^{\pi})+Y_{\nu_1,\nu_2}(i,J^{\pi})) \times \\ \\
    (u_{\nu_1}v_{\nu_2}+u_{\nu_2}v_{\nu_1})  
       \langle \nu_1||i^JY_{J}||\nu_2 \rangle\varphi_{\nu_1}(r)\varphi_{\nu_2}(r).\\
\end{array}
\label{eqn:td-density}
\end{equation}
The vertex $g$ is given by 
\begin{eqnarray}
& g_{\nu m J^{\pi}Mi}^{\nu' m'} = \sum_{L=J-1}^{J+1}
i^{l-l'}
\langle j'm'
| (i)^L[Y_L\times\sigma]_{JM}|j m\rangle \times \nonumber \\
\\ & \int d r\varphi_{\nu'} 
[(G_0+G_0')\delta\rho_{J^{\pi}Ln}^{i}+(G_0-G_0')\delta\rho_{J^{\pi}Lp}^{i}]
\varphi_{\nu}, \nonumber \\
\label{eqn:4}
\end{eqnarray}
where $G_0,G_0'$ are the generalized Landau-Migdal parameters  
controlling the isoscalar and isovector spin-dependent channels,
while $\delta\rho^i_{J^{\pi}Ln}$ and $\delta\rho^i_{J^{\pi}Lp}$ are respectively 
the neutron and proton
contributions to the transition densities 
\begin{equation}
\begin{array}{c}
\delta\rho_{J^{\pi}L}^{i}(r)=\frac{1}{\sqrt{2J+1}}
\sum_{\nu_1,\nu_2}(X_{\nu_1,\nu_2}(i,J^{\pi})-Y_{\nu_1,\nu_2}(i,J^{\pi})) \\
\\
\times (u_{\nu_1}v_{\nu_2}+u_{\nu_2}v_{\nu_1}) \times  
\langle \nu_1||i^L[Y_L\times\sigma]_J|| \nu_2\rangle\varphi_{\nu_1}(r)
\varphi_{\nu_2}(r). \\
\end{array}
\label{eqn:td-spin}
\end{equation}
The values of the Landau-Migdal parameters  associated with the SLy4 interaction
are shown in Fig. \ref{Landau}.
We observe that  only the vertices $g$, associated with spin-dependent part of the residual interaction
can contribute in the case of non-natural parity phonons (for which $J=L+1$ or $J=L-1$), 
while both $f$ and $g$   can contribute in the case of natural-parity phonons (for which $J=L$).
We have included phonons of both parities having energy up to 30 MeV,
associated with multipolarities from $J = $0 up to   $J = 5$.
We have verified that the results are essentially the same including multipolarities up to  $J = 8$.
This is in keeping with the fact that low-lying vibrations tend to lose their collective character,
when the associated wavelength  becomes of the order of the interparticle
distance, or smaller  than it.  
The calculation of the matrix elements of the induced interaction is 
then the same performed in ref.  \cite{Gori_Sn120}, except for the fact that there the SkM* interaction
was used, instead of the SLy4 (the influence of  $0^+,0^-$ and $1^-$ multipolarities,
which were not included in \cite{Gori_Sn120}, is negligible) . 
The main difference between the two interactions lies in the 
value of the effective mass, which is higher in the SkM* case, corresponding to a
higher level density close to the Fermi energy and therefore leading to larger pairing gaps. 

We remark that only the results obtained making use of the total interaction
$v_{Arg+ind}$ have physical meaning and  should be compared with experiment. 
However, in order to 
better understand the properties
of the total interaction and to make contact with the literature we shall also study the
Argonne and the induced interaction separately. In these two cases the matrix elements will not be multiplied by $Z$.

The diagonal matrix elements $\Delta_{nnlj}$ 
of the state-dependent pairing gap  obtained solving 
the HFB equations  with the matrix elements $v_{Arg + ind}$ (cf. Eq.(\ref{Int:arg+indotta})) are shown in 
Fig.~\ref{Delta:matrix:el} (squares). 
We plot the results for  single-particle states with energy less than 
100 MeV but we note that to reach convergence  within 100 keV for the pairing gap 
calculated with the Argonne interaction 
we have to include   single-particle levels 
with energy up to 800 MeV in the HFB equations. 
For clarity, here and in 
following figures, the matrix elements for $e_{nlj}>0$ have been averaged over intervals
of 3 MeV width.
The value of the pairing gap averaged over the five 
single-particle states 
close   to the Fermi energy (taking into account their degeneracy, i.e. 
$\Delta_F \equiv \sum_{\nu} (2j_{\nu}+1)\Delta_{nnlj}/\sum_{\nu}(2j_{\nu}+1)$,
where the sum extends over  $\nu = 3s_{1/2}, 2d_{5/2},2d_{3/2},1g_{7/2}$ and $1h_{11/2}$)    
is equal to $\Delta_F = 1.47 $ MeV, very close to the value  
derived from the experimental binding energies through the usual 
three-point formula. We also show  by diamonds the values of $\Delta_{nnlj}$ 
obtained with the Argonne pairing  interaction alone, corresponding to 
a value $\Delta_F = 1.04$ MeV. For large values of $e_{nlj}$, they assume small negative values due to the presence  of the strong repulsive core~\cite{Baldo90}. 
The  state dependent gaps obtained solving the HFB equations 
including only the induced interaction 
are also shown by triangles in Fig.~\ref{Delta:matrix:el}: the gap 
is concentrated close to the Fermi energy, and  $\Delta_F$= 1.11 MeV.
Negative values of the pairing gap associated with deep-lying levels 
are caused by the spin-dependent part of the induced interaction, associated with the Landau parameters $G_0$ and $G'_0$  which has a repulsive
character, as discussed in ref. \cite{Gori_Sn120}. 
This can be seen looking at Fig. \ref{Fig_app2}(a) in Appendix B, where we report the same kind of calculations
shown in Fig. \ref{Delta:matrix:el}, but including only the spin-independent part of the induced interaction
(that is, putting the Landau parameters $G_0$ and $G'_0$ equal to zero): 
in this case, the induced interaction alone leads to $\Delta_F = 1.88 $ MeV,
while adding the bare interaction (together with the $Z-$factor, 
cf. Eq. (\ref{Int:arg+indotta}))
one obtains $\Delta_F$ =  2.12  MeV.
It is difficult to determine
the spin-dependent  part of the particle-hole  interaction, and the balance between    
attraction and repulsion is rather dependent on the adopted parametrization. However,
the main factors  determining the induced interaction in finite nuclei  are 
the pronounced collective character of the surface modes, as well as the dominance of
neutron-proton interaction over neutron-neutron interaction: they determine
its overall attractive character, in contrast with the case of uniform neutron matter
\cite{Gori_Sn120}.
Therefore, while the absolute value of the pairing gap could be somewhat different employing another interaction,
we expect that the main trends of the spatial dependence discussed below would not be affected.

\begin{figure}[!t]
\begin{center}
\includegraphics*[width=6.3cm,angle=-90]{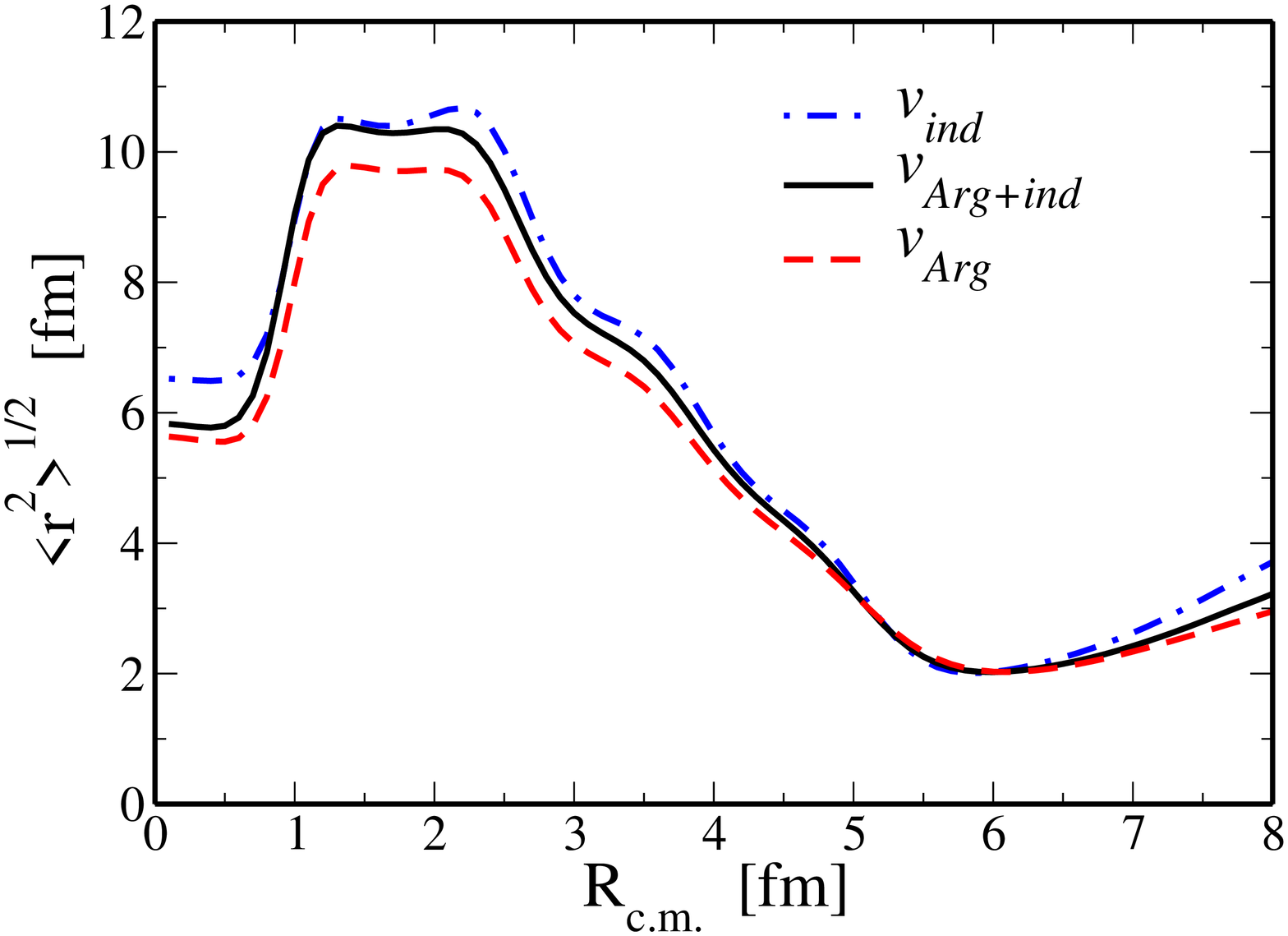}
\end{center}
\caption{\footnotesize{Root mean square radius of the Cooper pair as a function
of the position of the center of mass, obtained with the Argonne+induced interaction $v_{Arg + ind}$ (solid line), 
the Argonne interaction $v_{Arg}$ (dashed line), the induced interaction $v_{ind}$
(dash-dotted line).}}
\label{Fig:phir12}
\end{figure}

We also notice that the induced interaction is dominated by the contribution
of isoscalar modes, while isovector modes reduced the gap slightly.
While $T$ is not a good quantum number, we have in fact  found that including only the modes
that have a dominant $T=0$ character the pairing gap produced by the induced interaction 
is increased from 1.11 to 1.30 MeV, 
while including only modes with dominant $T=1$ character we do not find any pairing gap.

For an interaction that depends only on the relative coordinate $r_{12}$, like
the Argonne interaction, the  pairing
field $\Delta(\vec r_1, \vec r_2)$ is directly related to the Cooper
pair wavefunction introduced in Eqs. (\ref{a:density:r}) and (\ref{2:wave}):
\begin{equation}\label{Delta:r:2}
 \Delta(\vec{r}_1,\vec{r}_2)=-v(r_{12})\Phi^{S=0}(\vec{r}_1,\vec{r}_2).
\end{equation}
We note that the matrix elements  of the induced
interaction (cf. Eq. (\ref{VIND}))
depend on the energies of the single-particle states through
the energy denominators, and  one cannot directly use Eq. (\ref{Delta:r:2}) to obtain a corresponding 
pairing field. We shall instead use the fact that either one of  the two bases 
${\psi_{nn'lj^{\uparrow}}}$, with $j^{\uparrow}= |l+1/2|, l=0,1,...$ 
or ${\psi_{nn'lj^{\downarrow}}}$, with  
and $j^{\downarrow}= |l-1/2|, l=0,1,2,...$ is a complete basis 
in the ($J=0,S=0$) subspace,
constructing the associated pairing fields $\Delta^{\uparrow}$
and $\Delta^{\downarrow}$, as 
\begin{equation}\label{Delta:r}
 \Delta^{\uparrow} (\vec{r}_1,\vec{r}_2)=\sum_{nn'lj^{\uparrow}} (2l+1)
 \Delta_{nn'lj^{\uparrow}}\psi_{nn'lj^{\uparrow}}^{S=0} (\vec{r}_1,\vec{r}_2),
\end{equation}
and similarly for $\Delta^{\downarrow}$,
where  the 
factor $(2l+1)$ is a normalization factor associated with the Legendre polynomials.
It turns out that there  is some dependence on which of the two basis is used.
This is due to the structure of the denominators in Eq. (6)
and to the effect of the spin-orbit interaction. 
For simplicity we shall limit ourselves in
the following to the pairing field
obtained taking the average  of the two expansions:
\begin{equation}\label{Delta_mediato}
\Delta_{ind} = \frac{\Delta^{\uparrow}+ \Delta^{\downarrow}}{2}. 
\end{equation}
 We shall show that this leads to a local
expression for the pairing field, which reproduces rather well  the
quasiparticle energies and the pairing energies obtained solving the original HFB
equations (\ref{HFB}).

\vskip 0.5cm

\noindent We shall now study the Cooper wavefunction
and the pairing field associated with the bare Argonne interaction
and the pairing induced interaction. 
In Fig.~\ref{Phi:Rr} we show  the Cooper pair wave-function $\Phi^{S=0}$ 
for fixed values of  $R_{c.m.}$ (the center of mass
of the pair), as a function of the relative distance $r_{12}$. 
The wavefunction also depends weakly
on the value  of the angle $\theta_p$ between $\vec R_{c.m.}$ and $\vec r_{12}$, 
and we show the result obtained after an angular average.
At small values of the relative distance, $r_{12} < $~1~fm, the  
strong repulsive core present in the Argonne
interaction prevents the two neutrons from staying close to each other,
producing a hole in the wavefunction (see Fig. \ref{Phi:Rr}(b)). For larger values of $r_{12}$ the 
wavefunctions are rather similar (see Fig \ref{Phi:Rr}(a), \ref{Phi:Rr}(b)), as can also be seen in Fig. \ref{Fig:phir12}
where we show the root mean square radius $\langle r_{12}^2\rangle^{1/2}$, as a function of 
the position of the center  of mass \cite{Pillet}. In fact, $\Phi^{S=0}$ is dominated 
by the spatial dependence  of the single-particle 
wavefunctions \cite{Sandulescu_PRC05}. 
One can remark that, due to the finite size of the nucleus,  which limits
the phase space available for the formation of  Cooper pairs,
the values of $\langle r_{12}^2\rangle^{1/2}$ are considerably smaller than 
the value of the coherence length $\xi$ in uniform neutron or nuclear matter
at the corresponding density. In fact, $\xi$  can be estimated 
from $\xi = \hbar^2 k_F/m^* \pi \Delta_F$ \cite{Degennes,PRC98_Barranco} , leading to
$\xi  \sim 19 $ fm inside the nucleus ($m^* = 0.7, k_F = 1.3 $ fm$^{-1}$, $\Delta_F$ = 1 MeV)
and to $\xi \sim  $ 6 fm on the surface ($m^* = 1, k_F = 0.9 $ fm$^{-1}$, $\Delta_F$ = 2 MeV).

In Fig.~\ref{Delta:Rr} we show the structure of the pairing field 
as a function of the relative  distance for various values 
of the center of mass coordinate $R_{c.m.}$, averaged on the angle $\theta_p$.
In the case of the Argonne interaction, 
the pairing field is obtained from Eq. (\ref{Delta:r:2}),  while in the
case of the induced interaction it is obtained  
from~Eq.(\ref{Delta_mediato}), as discussed above.
The repulsive core produces the large negative quantities  
at small values of $r_{12}$ observed in Fig.~\ref{Delta:Rr}(b),
while the attractive part prevails for $r_{12} >1$~fm. 
The  induced gap  (Fig.~\ref{Delta:Rr}(a)) is strongly peaked around $R_{c.m.} \approx $ 6 fm, 
in keeping with the fact that it receives the main contribution
from the low-lying collective modes, whose transition density is concentrated
on the surface of the nucleus \cite{Gori_Sn120,PRC05_Barranco}.
The negative values for small $R_{c.m.}$ and $r_{12}$ are due to the 
repulsive, spin-dependent part of the  interaction (cf. the corresponding 
Fig. \ref{Fig_app2}(b) 
where this part has been left out). 
We note the different energy scale in Figs. \ref{Delta:Rr}(a) and \ref{Delta:Rr}(b).
In fact, the presence of the  repulsive core in the bare interaction 
makes it difficult to  assess the relevance of the induced interaction.
It is more convenient to consider the dependence of the two interactions
on relative momentum $k$, because the effect of the repulsive core is then 
restricted to high values of $k$, for which the induced interaction plays no role.

\section{Momentum dependence of the pairing field and its local approximation}\label{sec:momento}

\begin{figure*}[!h]
\begin{center}
\includegraphics[width=6cm,angle=-90]{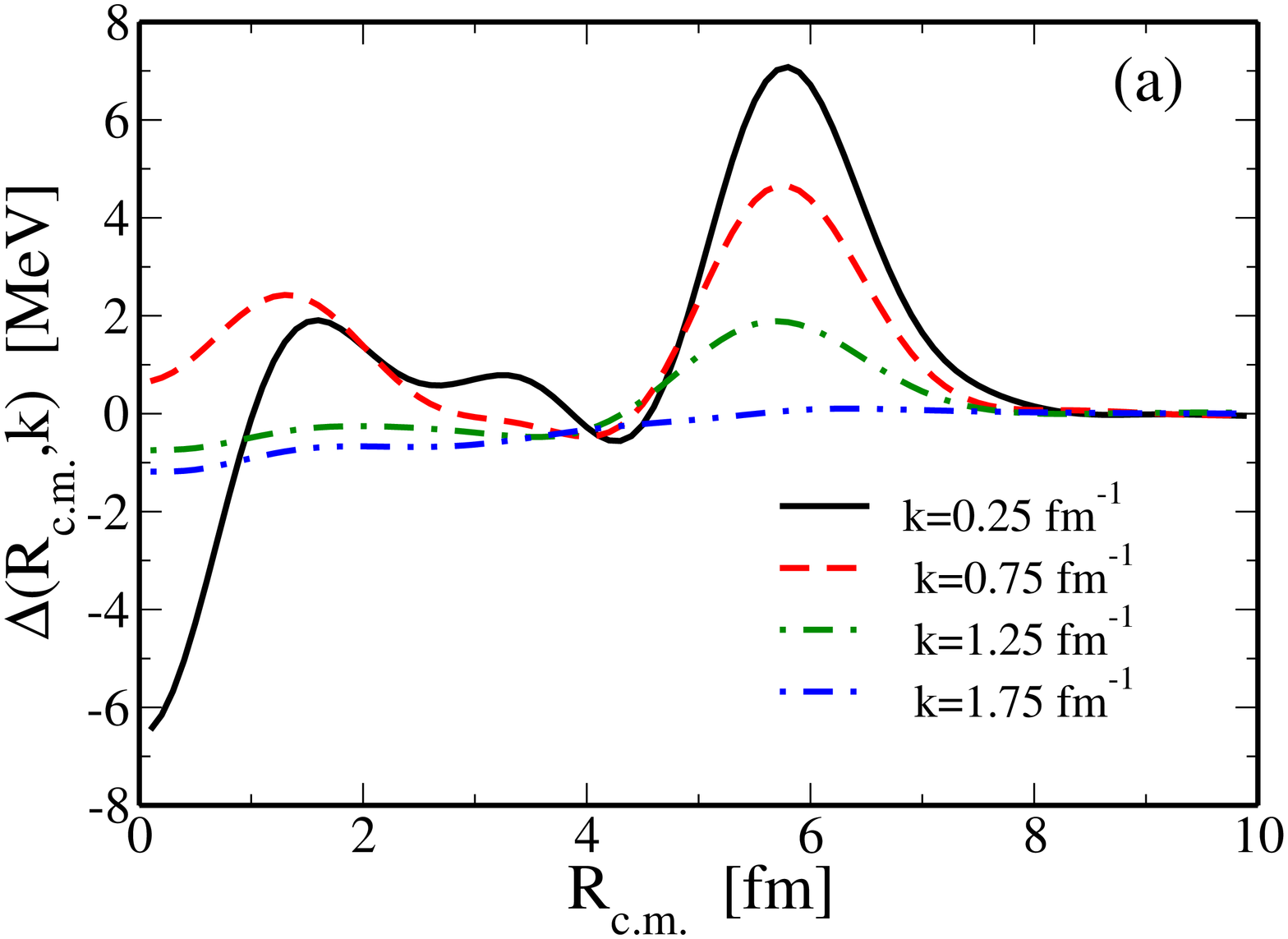}
\hspace{0.05\textwidth}
\includegraphics[width=6cm,angle=-90]{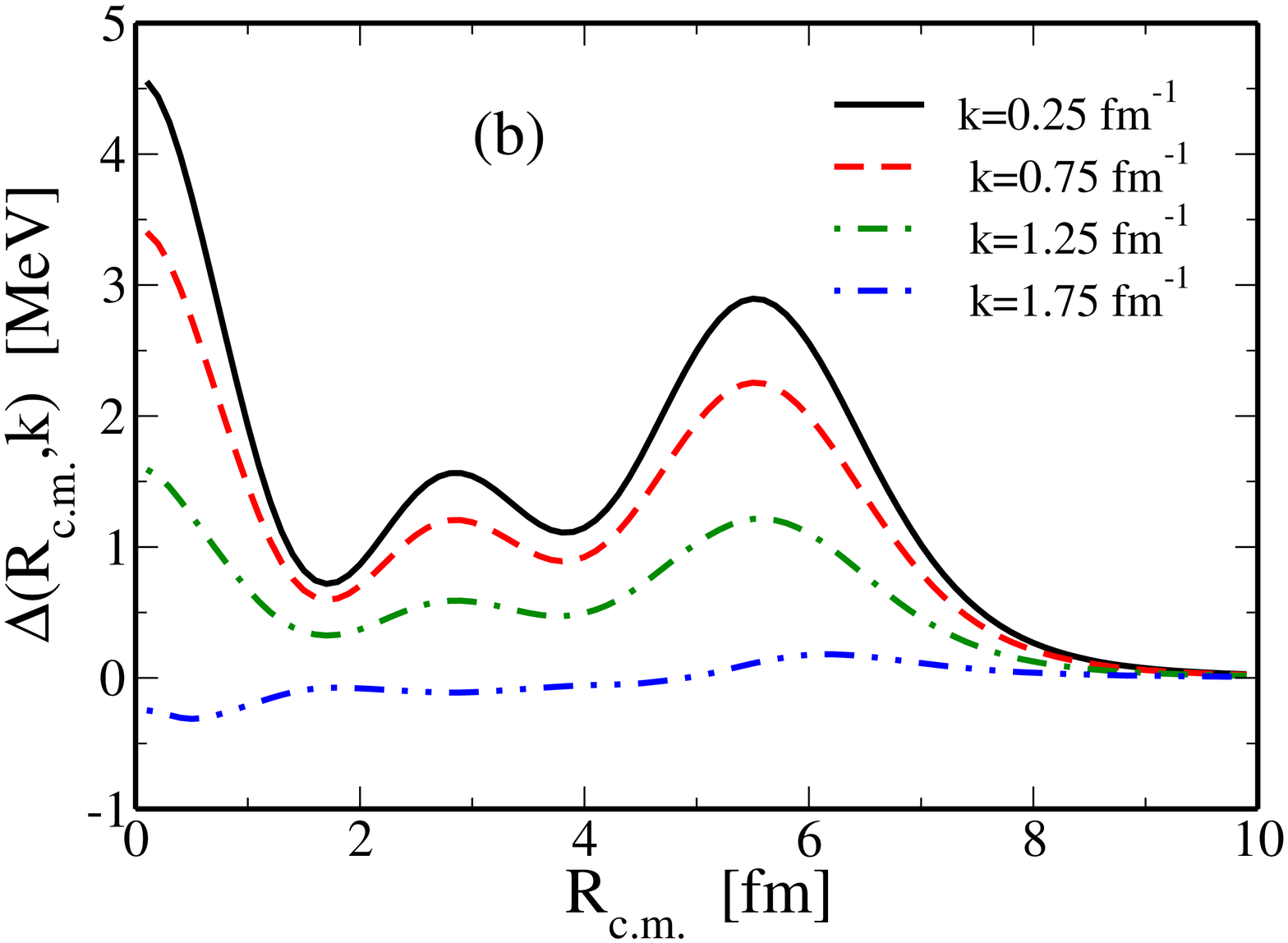}
\end{center}
\caption{\footnotesize{Pairing field (\ref{Delta:Fourier}) 
as a function of the position of the center of mass
for different values of the relative 
momentum $k$, for the Argonne plus induced interaction $v_{Arg + ind}$ (a) and 
for the 
Argonne interaction $v_{Arg}$ (b).}}
\label{Delta:Kr:A+i:A}
\end{figure*}

In this section we study the momentum dependence of the pairing field, by
taking the Fourier transform of $\Delta(\vec r_1, \vec r_2)$ 
with respect to  
the relative distance $\vec{r}_{12}$:
\begin{equation}\label{Delta:Fourier}
 \Delta(\vec{R}_{c.m.},\vec{k})=\frac{1}{(2\pi)^3}\int d^3r_{12}e^{i\vec{k}\cdot \vec{r}_{12}}
 \Delta(\vec{R}_{c.m.},\vec{r}_{12}).
\end{equation}

\noindent We then average on the angle between  $\vec{R}_{c.m.}$ 
and the relative momentum $\vec{k}$,
and obtain a function $\Delta(R_{c.m.},k)$ that depends only 
on the moduli of these two vectors.
In Fig.~\ref{Delta:Kr:A+i:A} we plot $\Delta(R_{c.m.},k)$ for $v_{Arg}$ and for the total interaction $v_{Arg + ind}$.
For the bare interaction, the behavior at small values of $R_{c.m.}$ 
is dominated by 
the $3 s_{1/2}$ orbit, while the negative values of $\Delta$ 
at high values of $k$ are due to the repulsive core~\cite{Baldo90}.
Adding the induced interaction clearly has a strong effect on the pairing
field for values of $k$ lower than about 1 fm$^{-1}$, enhancing the gap 
in the surface region and reducing it inside the volume of the nucleus.

\begin{figure}[htb]
\begin{center}
\includegraphics[width=6cm,angle=-90]{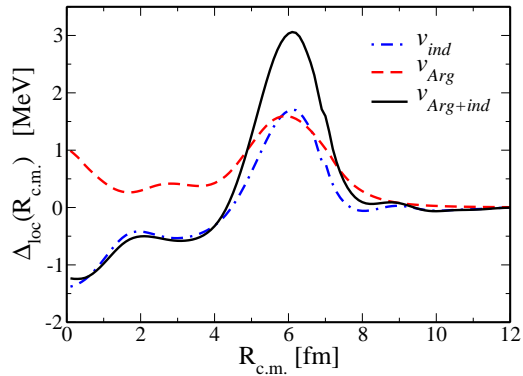}
\end{center}
\caption{\footnotesize{Pairing field obtained with the semiclassical 
approximation (cf. Eq. (\ref{Delta:approx})) for the three different pairing interactions: 
Argonne plus induced $v_{Arg + ind}$ (solid line), Argonne $v_{Arg}$ (dashed line), 
induced $v_{ind}$ (dash-dotted line).}}
\label{Delta:Fermi}
\end{figure}

One can obtain  a local approximation to the pairing field, 
through a simple Thomas-Fermi approximation~\cite{Ring&Schuck,PRC98_Barranco} 
writing $\Delta_{loc} (R_{c.m.}) \equiv \Delta(R_{c.m.},k_F(R_{c.m.}))$, where 
the local Fermi momentum is given by 
\begin{equation}\label{Fermi:mom}
 \hbar^2k_{F}^2(R_{c.m.})=2m^*(R_{c.m.})(e_F-U(R_{c.m.}))),
\end{equation}
\noindent where $U(R_{c.m.})$ is the HF potential, and $m^*(R_{c.m.})$ 
is the effective mass \cite{Vautherin_PRC5}  
associated with the SLy4 interaction.
The expression (\ref{Fermi:mom}) is only valid in the classically 
allowed  region where $e_F-U(R_{c.m.})>0$ 
(in the present case the turning point lies at $R_{t} = $ 7.1 fm). 
We shall extend our definition into the classically 
forbidden region,   using the Fourier transform at zero momentum:
\begin{eqnarray}
\Delta_{loc}(R_{c.m.})_{ext} &\equiv& \Delta(R_{c.m.},k=0) \nonumber \\
                           &=&\frac{1}{(2 \pi)^3} \int d^3r_{12} \Delta(\vec{R}_{c.m.},\vec{r}_{12}).
\label{Delta:loc}
\end{eqnarray}
This is equivalent to using a local momentum associated with an energy 
slightly larger than $e_F$. 
In this way the pairing field 
\begin{eqnarray}\label{Delta:approx}
\Delta_{loc}(R_{c.m.})=
\begin{cases}
     \Delta(R_{c.m.},k_{F}(R_{c.m.})) & \text{$R_{c.m.}\leq R_{t}$};\\
                       \Delta_{loc}(R_{c.m.})_{ext} & \text{$R_{c.m.}>R_{t}$}.\\
\end{cases}
\end{eqnarray}
\noindent is continuous, and we have found that the first derivatives also match to a good
approximation. 
The resulting local pairing fields are plotted in
 Fig.~\ref{Delta:Fermi}. 
The pairing field associated with the Argonne interaction is 
rather surface peaked, going from 1.5 MeV at the surface 
to 0.5 MeV in the interior. Adding  the induced interaction 
reinforces this
surface character, leading to 
a large  peak at the surface of about 3 MeV. 
The negative values of the gap
in the interior of the nucleus are caused by the spin-dependent part of the 
induced interation, as can be checked, comparing with Fig. \ref{Fig_app2}(f) in Appendix B,
obtained including only the spin-independent part of the interaction: in that 
case the pairing gap essentially vanishes inside the nuclear volume, while it reaches 
a value of about 4  MeV on the surface.

\begin{figure}[htb]
\begin{center}
\includegraphics[width=6cm,angle=-90]{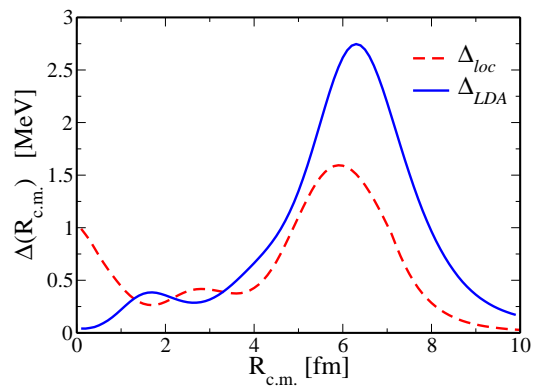}
\end{center}
\caption{\footnotesize{The local pairing gap calculated from Eq.(\ref{Delta:approx}))  for the Argonne interaction, 
already shown in Fig. \ref{Delta:Fermi} (dashed line), is compared to the gap obtaining using the simple
LDA approximation (solid line) }}
\label{LDAsemi}
\end{figure}

The local pairing field  $\Delta_{loc}(R_{c.m.})$ can  
be used as the pairing potential  
in the HFB equations for a zero-range potential written in
coordinate space \cite{Hamamoto_PRC68}:
\begin{eqnarray}\label{HFB-r}
&&\left(\frac{d^2}{dR^2}-\frac{l(l+1)}{R^2}+\frac{2m^*}{\hbar^2}
[e_F+E_{qp}-U(R)]\right)u_{lj}(R) \nonumber \\
&&+\frac{2m^*}{\hbar^2}\frac{d}{dR}\left(\frac{\hbar^2}{2m^*}\right)
\frac{d}{dR}u_{lj}(R)-\frac{2m^*}{\hbar^2}\Delta_{loc}(R)v_{lj}(R)=0,\nonumber \\
&&\left(\frac{d^2}{dR^2}-\frac{l(l+1)}{r^2}+\frac{2m^*}
{\hbar^2}[e_F-E_{qp}-U(R)]\right)v_{lj}(R) \nonumber \\
&&+\frac{2m^*}{\hbar^2}\frac{d}{dR}\left(\frac{\hbar^2}{2m^*}\right)\frac{d}{dR}
v_{lj}(R)+\frac{2m^*}{\hbar^2}\Delta_{loc}(R)u_{lj}(R)=0. \nonumber \\
&&
\end{eqnarray}
\noindent To test the reliability of the semiclassical $\Delta_{loc}(R_{c.m.})$ we have compared the quasi-particle energies 
and the occupation probabilities obtained solving the self-consistent HFB equations 
(cf. Eq.~(\ref{HFB})) using the full potentials, with the solution of 
Eq. (\ref{HFB-r}) obtained using the local potential.
The results are collected in Table \ref{tab:Eqp}. The overall agreement  
is rather good: most quasiparticle
energies are reproduced within 200 keV and the occupation probabilities 
larger than 0.1 are reproduced within 15\%. 
The local approximation introduced above, based on the results  obtaiend in the
microscopic HFB calculation, leads to pairing gaps which are 
rather different from those obtained from the simplest Local Density Approximation, 
which does not take into
account  proximity effects associated with the  nuclear surface and the fact that the nuclear radius is smaller
than the coherence length in uniform matter. This can be seen in Fig. \ref{LDAsemi},
where we compare the local pairing gap $\Delta_{loc}$ associated with the Argonne
interaction, with the function $\Delta_{LDA}(R_{c.m.}) = \Delta_F^{n.m.}(\rho_n(R_{c.m.}))$,
where $\Delta_F^{n.m.}$ is the pairing gap calculated at the Fermi energy in uniform neutron matter,
for a density equal to the neutron density at a distance $R_{c.m.}$ from the center of the nucleus,
and using the local value of the effective mass. The LDA overestimates the difference between the
pairing gap on the surface and in the interior of the nucleus, an effect already observed in
the case of the inner crust in neutron stars \cite{Pizzochero}.  
\vspace{1cm}

\begin{table*}
\begin{center}
\begin{tabular}{c|cccc|cccc|cccc}
\hline
\hline
&\multicolumn{4}{c|}{$v_{Arg + ind}$} & \multicolumn{4}{c|}{$v_{Arg}$} & \multicolumn{4}{c}{$v_{ind}$} \\
$l\;\;2j$ &$E_{qp}^{full}$ &  $v^2_{can}$ & $E_{qp}^{loc}$ & $v^2_{loc}$ &$E_{qp}^{full}$ & $v^2_{can}$ & $E_{qp}^{loc}$ 
& $v^2_{loc}$ & $E_{qp}^{full}$ & $v^2_{can}$ & $E_{qp}^{loc}$ & $v^2_{loc}$\\
\hline
\hline
0  1 & 1.92 & 0.76 & 1.35& 0.86 & 1.41&  0.85& 1.31 & 0.86 & 1.30& 0.78& 0.91& 0.95\\
2  3 &1.49 & 0.66 &1.30 &0.69  & 1.18&  0.72& 1.08 & 0.74 & 0.87& 0.68& 0.67& 0.77\\
2  5 &3.76 & 0.93 &3.48 &0.97  &3.46&  0.98& 3.42 & 0.98 & 3.51& 0.94& 3.19& 0.99\\
4  7 &2.21 & 0.94& 2.32& 0.93 & 2.33&  0.94& 2.27 & 0.95  & 1.87& 0.99& 1.98& 0.98\\
5 11 & 1.89& 0.23& 1.88& 0.25  &  1.37&  0.15& 1.48 & 0.17  & 1.84& 0.18& 1.38& 0.10\\
\hline
\hline
\end{tabular}
\caption{\footnotesize{ The lowest quasiparticle energies, expressed in MeV, associated with the quantum numbers $(l,2j)$ 
obtained solving the HFB equations Eq.~(\ref{HFB})  with the
Argonne+induced, Argonne and induced interactions are indicated with $E_{qp}^{full}$; also listed are the occupation
probabilities $v^2_{can}$ obtained in the canonical basis. They are compared with the quasiparticle energies 
$E_{qp}^{loc}$ and occupation probabilities $v^{2}_{loc}$ obtained solving the HFB equations 
(Eq. (\ref{HFB-r}))  in coordinate space
with the local pairing potentials shown in  Fig.~\ref{Delta:Fermi} 
and discussed in the text.}}\label{tab:Eqp}
\end{center}
\end{table*}


\clearpage

\section{LDA parametrization of the pairing interaction }\label{DDDI_fit}

\subsection{Density dependent, zero-range parametrization} 

\begin{figure*}[!t]
\begin{center}
\includegraphics*[width=6.3cm,angle=-90]{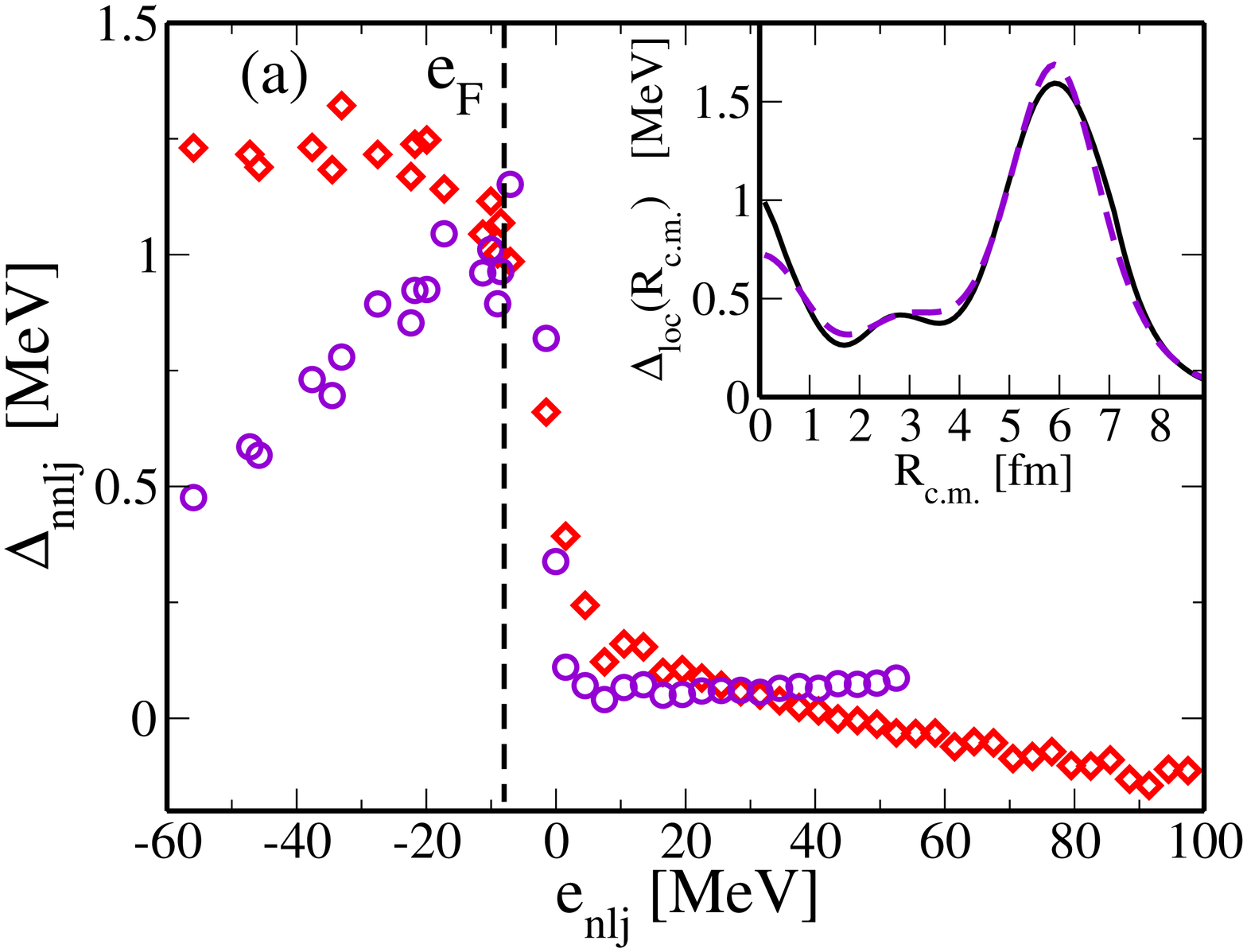}
\includegraphics*[width=6.3cm,angle=-90]{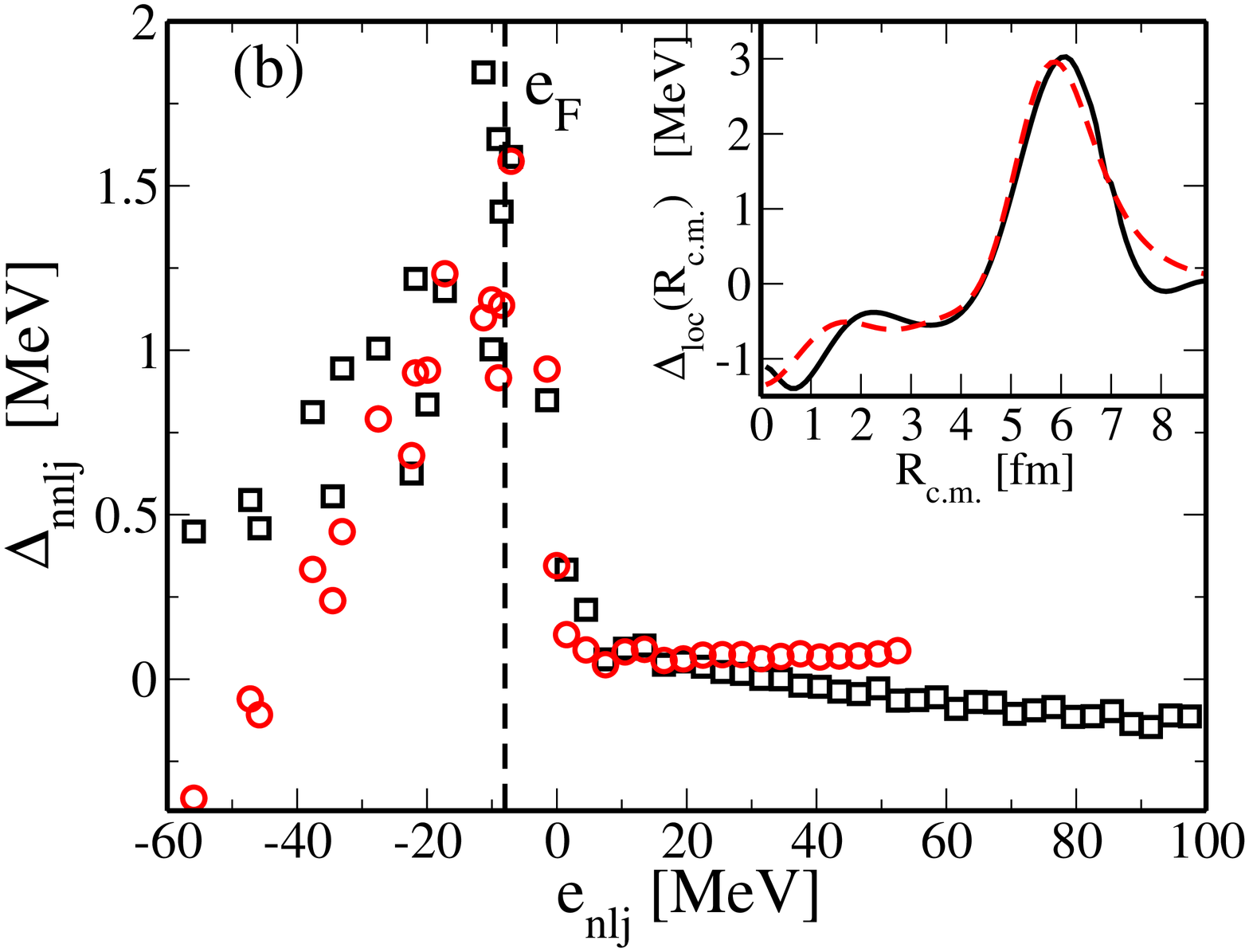}
\end{center}
\caption{\footnotesize{
(a) The diagonal matrix elements of the pairing gap associated with the Argonne 
interaction $v_{Arg}$ (diamonds, already shown in  Fig. \ref{Delta:matrix:el}) 
 are compared with those associated with the DDDI, zero-range
interaction  with the parameters $\alpha=0.66$, $\eta = 0.84$ (circles). The 
semiclassical pairing gaps associated with the Argonne interaction (solid line)
and with the zero-range interaction (dashed line) are shown in the insert.
The diagonal matrix elements of the pairing gap  associated with  Argonne+induced interaction $v_{Arg + ind}$ (squares, already shown in Fig. \ref{Delta:matrix:el}) are compared with the DDDI interaction with 
the parameters $\alpha=2.0$, $\eta = 1.32$ (circles, cf. Table~\ref{Tab:DDDI}).
The 
semiclassical pairing gaps associated with the induced interaction (solid line)
and with the zero-range interaction (dashed line) are shown in the insert. }}
\label{fig:para1}
\end{figure*}

\begin{figure*}[!t]
\begin{center}
\includegraphics*[width=6.3cm,angle=-90]{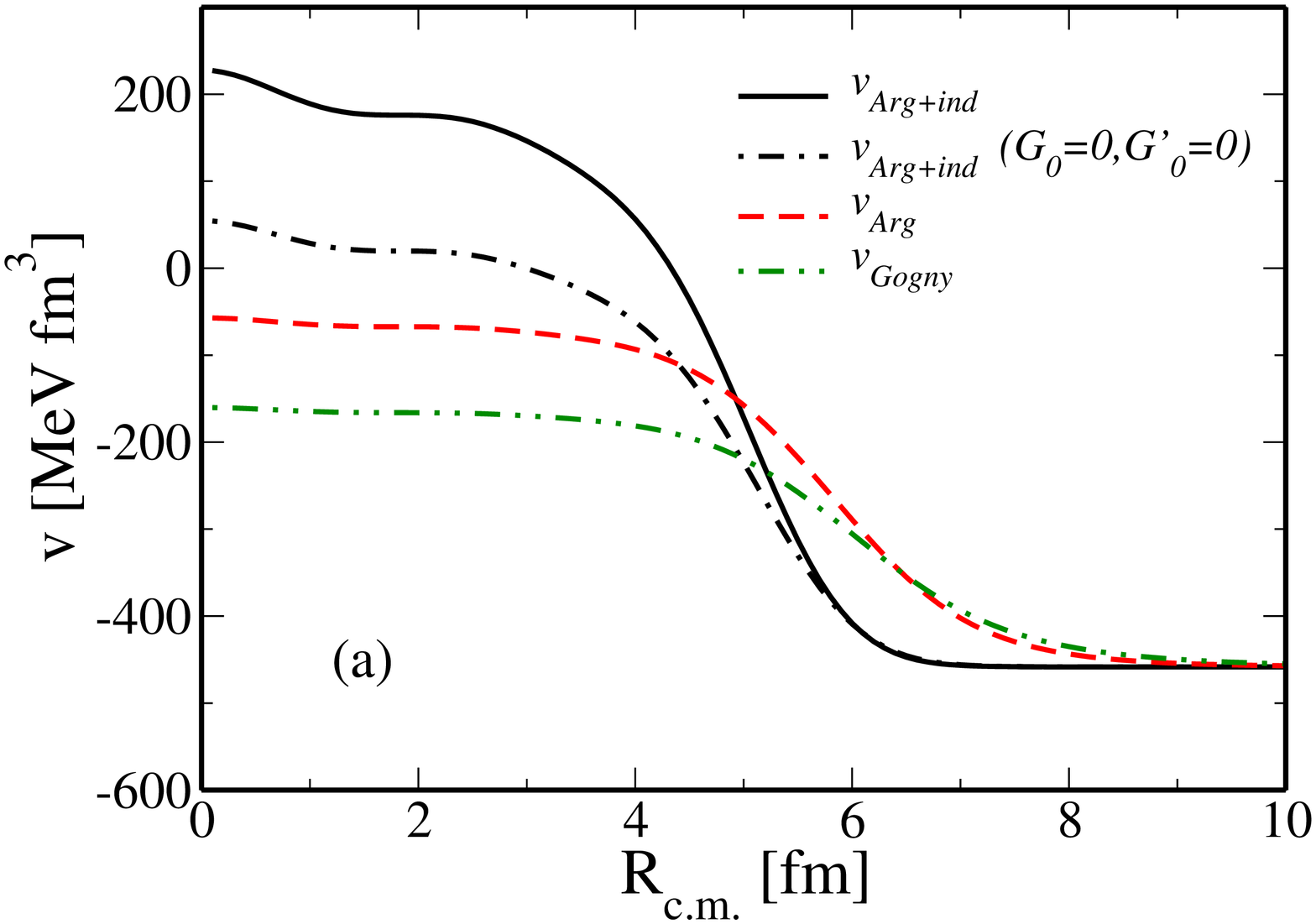}
\includegraphics*[width=6.3cm,angle=-90]{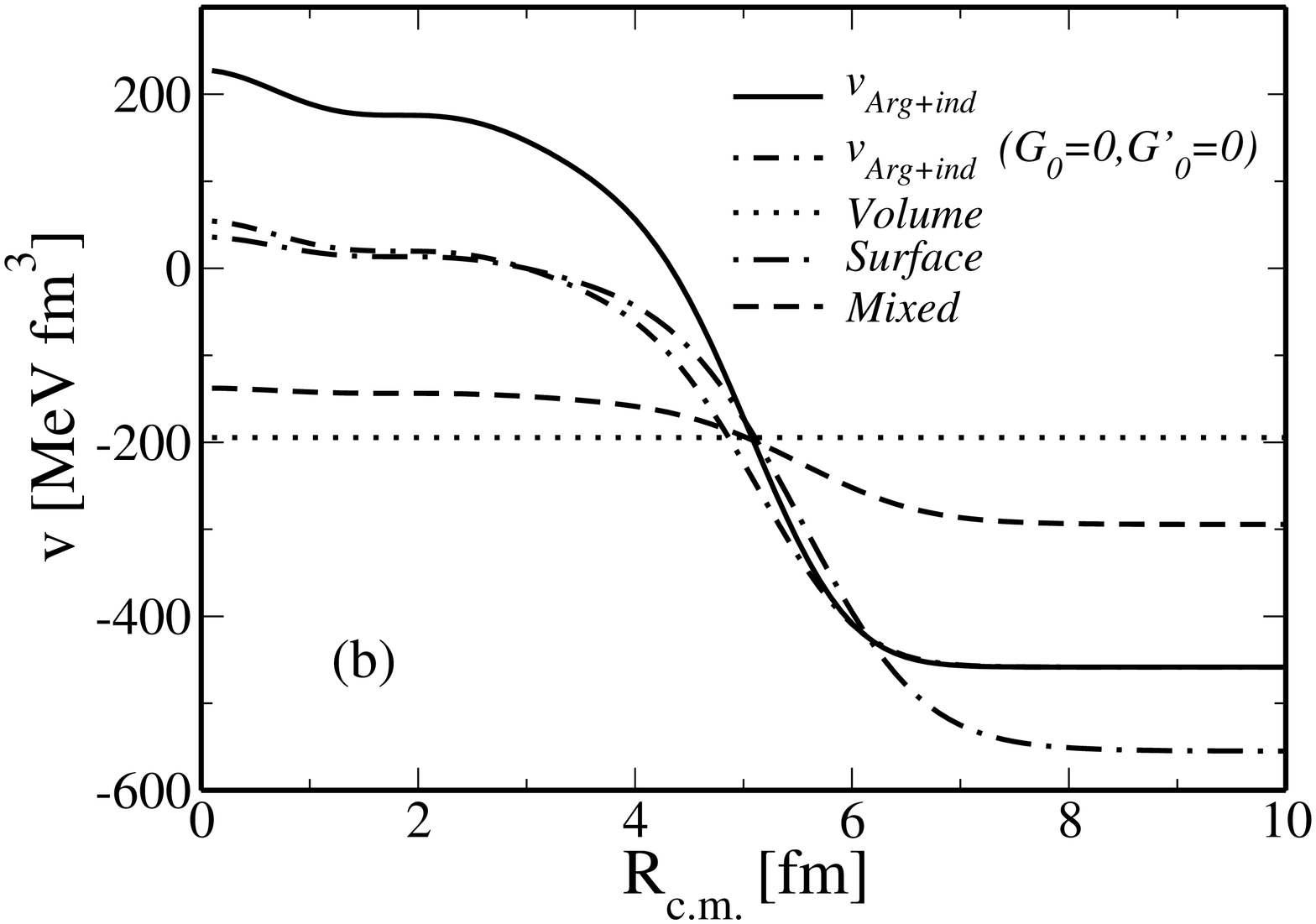}
\end{center}
\caption{\footnotesize{(a) 
Spatial dependence of the different local pairing interactions introduced  in this work
to simulate the local pairing gaps (cf. Eq. (\ref{Delta:approx})) 
obtained with the corresponding microscopic, non local interactions:
bare+induced interaction $v_{Arg+ind}$ (corresponding to the
parameters $\alpha=2.0,\eta=1.32$,  cf. Table \ref{Tab:DDDI}); 
bare+induced interaction neglecting the spin-dependent part
($v_{Arg+ind}, G_0=0, G'0=0,$ ($\alpha=1.79,\eta=1.0$,  cf. Table \ref{Tab:Table_app2});
bare $v_{14}$ interaction $v_{Arg}$ ($\alpha=0.66,\eta=0.84$,  cf. Table \ref{Tab:DDDI}); 
Gogny  interaction $v_{Gogny}$ ($\alpha=0.51,\eta=0.63$,  
cf. Table \ref{Tab:Table_app1} ). 
(b) The spatial dependence of the bare+induced interaction with and without 
the spin-dependent part of the induced interaction, already shown in (a), are compared 
with the volume, surface and mixed interaction of ref. \cite{Doba_inter} (see text).}}
\label{fig:interaction}
\end{figure*}

\begin{table}[b!]
\begin{center}
\begin{tabular}{c|c|c|c|c|c|c}
\hline
\hline
Interaction                & $\alpha$ & $\eta$ & $\Delta_F^{full}$ & $\Delta_F^{\delta}$ & $E_{pair}^{full}$& $E_{pair}^{\delta}$\\
\hline
$v_{Arg }$               & 0.66  & 0.84 &  1.04 & 1.03 & -13.2 & -8.9\\
$v_{Arg + ind}$ & 2.0  & 1.32  &  1.47& 1.28 & -15.78 & -14.47\\
\hline
\hline
\end{tabular}
\end{center}
\caption{\footnotesize{Parameters of the DDDI, Eq. (\ref{DDDI}), 
producing pairing gaps which  fit the  
local semiclassical pairing fields obtained with the various interactions.
In the last two columns we compare the pairing gap at the Fermi energy and the pairing energies (in MeV)  obtained with the full calculation, $\Delta_F^{full}$ and 
$E_{pair}^{full}$, with the values obtained using the corresponding density dependent 
interaction, $\Delta_F^{\delta}$ and $E_{pair}^{\delta}$.}}
\label{Tab:DDDI}
\end{table}

The local pairing fields discussed in the previous section can be compared to those 
obtained by several authors, who employed a density-dependent pairing interaction (DDDI) of
the form \cite{Bertsch_AP91}-\cite{Matsuo_PRC06}:
\begin{equation}\label{DDDI}
v^{\delta}(\vec{r}_1,\vec{r}_2)=v_0\left[1-\eta\left(\frac{\rho\left(\frac{\vec{r}_1+\vec{r}_2}{2}\right)}{\rho_0}\right)^{\alpha}\right]\delta(\vec{r}_1-\vec{r}_2),
\end{equation}
 \noindent where $\rho_0$ is the nuclear saturation density and $v_0,\eta,\alpha$ are three parameters
 to be determined, together with the value of a cutoff energy in the single-particle
 energies $E_{cut}$, needed to solve the HFB equations
 with a zero-range interaction \cite{Bulgac_PRL02,Dobaczewski_PRC06}.
The parameter $v_0$ together with $E_{cut}$ defines the strength of the pairing interaction, 
while the other two parameters determine the shape of the pairing field.
For a given value of $E_{cut}$, the strength can be fixed at zero density so as to reproduce 
the neutron scattering length. We shall use the single-particle levels which lie 
up to 60 MeV above the Fermi energy, following ref.~\cite{Matsuo_PRC06},
and as a consequence  we shall put $v_0 =  -458.4 $ MeV fm$^{-3}$. 

The  parameters $\alpha$ and $\eta$ have been  determined  in previous works
either to reproduce 
experimental gaps or to reproduce the pairing gap at the Fermi energy 
obtained with a finite range interaction like Gogny or Argonne in uniform neutron matter.
In this section we want instead to determine the parameters of the DDDI
 from the condition that the spatial dependence of the associated gaps reproduces 
 that of the  local pairing fields 
determined in the previous section (cf. Eq. (\ref{Delta:approx})) .
We solve the HFB equations (cf. Eq.~(\ref{HFB})) 
for the pairing interaction (\ref{DDDI}).
We then fit the parameters $\eta$ and $\alpha$, minimizing the deviation  
between the form of the pairing gap obtained
with the DDDI interaction of Eq. (\ref{DDDI}) and  the form of the gap  
obtained with the  local potentials. 
The values of the parameters for the various interactions are reported in Table~\ref{Tab:DDDI}.
Interestingly, the values we obtain for the Argonne interaction are  
very close to those obtained by Matsuo 
for the bare interaction in uniform neutron matter~\cite{Matsuo_PRC06}. 
The values obtained for the Argonne+induced interaction 
correspond to a larger attraction in the surface region, in keeping with 
Fig.~\ref{Delta:Fermi}.
The  diagonal matrix elements of the pairing gap associated 
with $v_{Arg}$ and  $v_{Arg + ind}$, already shown in Fig.~\ref{Delta:matrix:el}, are compared
with the corresponding quantities obtained
using the zero-range interaction in Fig.~\ref{fig:para1}, where we also compare the spatial dependence of the local pairing gaps (see insets). 
One can notice that the 
zero-range interaction (DDDI) yields a larger value of the gap for the levels above 
the Fermi energy. Nevertheless,
we are able to reproduce the pairing energies associated with  
the Argonne+induced interaction
within an accuracy of about $10\%$ (cf. Table~\ref{Tab:DDDI}). 
The agreement is not so good in the case of the pure Argonne interaction. 
In this case we  could improve the agreement between 
the pairing energies modifying the parameters slightly. We have 
found that using the parameters $\alpha=0.7,\eta=0.8$ 
we can reproduce the pairing energy within 
an accuracy of better than $5\%$, worsening somewhat 
the reproduction of the spatial dependence of the gap.

In Fig. \ref{fig:interaction}(a) we compare the spatial dependence of the various local, 
density-dependent interactions, introduced above and also in Appendix A and B.
The  bare+induced interaction is considerably more attractive  
than the bare Argonne interaction  or the effective Gogny 
interaction for $R_{c.m.} \sim $ 6 fm. The effect of the spin-dependent part of the interaction, which 
produces a repulsive contribution in the nuclear interior is also clearly seen
in the figure. By construction, all the interactions tend to the value $v_0 = - 458.4$ MeV fm$^{-3}$  for large values of $R_{c.m.}$.  
In Fig.  \ref{fig:interaction}(b) we compare our
results for the bare+induced interactions (with and without the spin-dependent part)
with the three schematic  DDD interactions proposed in ref. \cite{Doba_inter}, where
the associated pairing gaps  have been compared with those extracted from the
experimental  odd-even mass differences. These interactions are of  the form 
(\ref{DDDI}) with $\alpha=1$, and with $\eta = 0$ (volume force), $\eta $ = 1 (surface
force)  and $\eta =  0.5$ (mixed force). 
The value of $v_0$ in this case has been obtained imposing that the average value  of the pairing  field weighted with the 
nuclear  density,
$\bar{\Delta} \equiv \int d^3r \Delta(r) \rho(r) $, be equal to 1.24 MeV (the cutoff  adopted in
\cite{Doba_inter} is slightly different from ours, and we have imposed the same 
condition within our space).  The definition of $\bar{\Delta}$  
gives more weight to the value of the pairing field in the interior, compared
to our definition of $\Delta_F$, which is based on the single-particle levels at
the Fermi energy, which are more localized on the nuclear surface.

\begin{figure*}[!t]
\begin{center}
\includegraphics*[width=6.3cm,angle=-90]{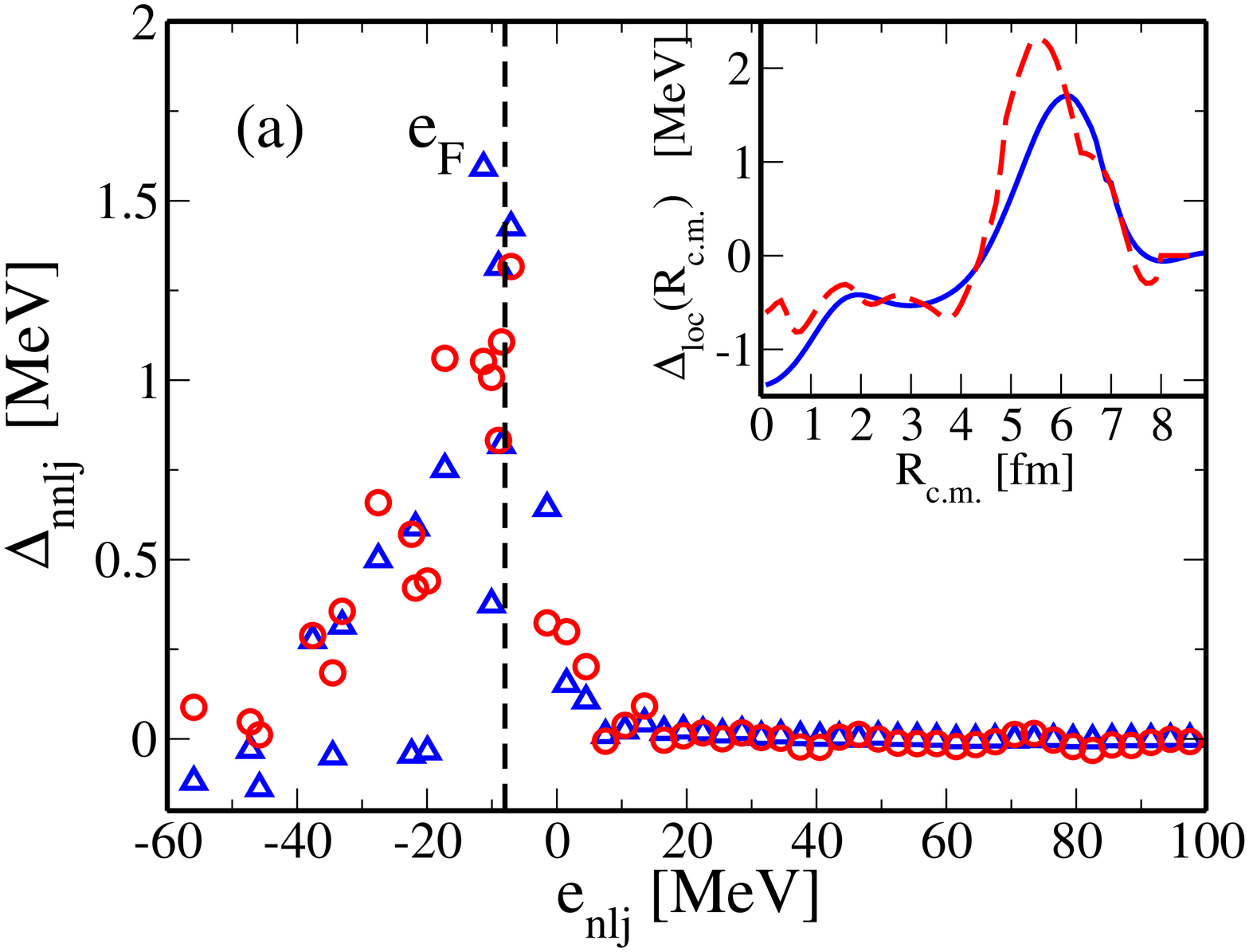}
\includegraphics*[width=6.3cm,angle=-90]{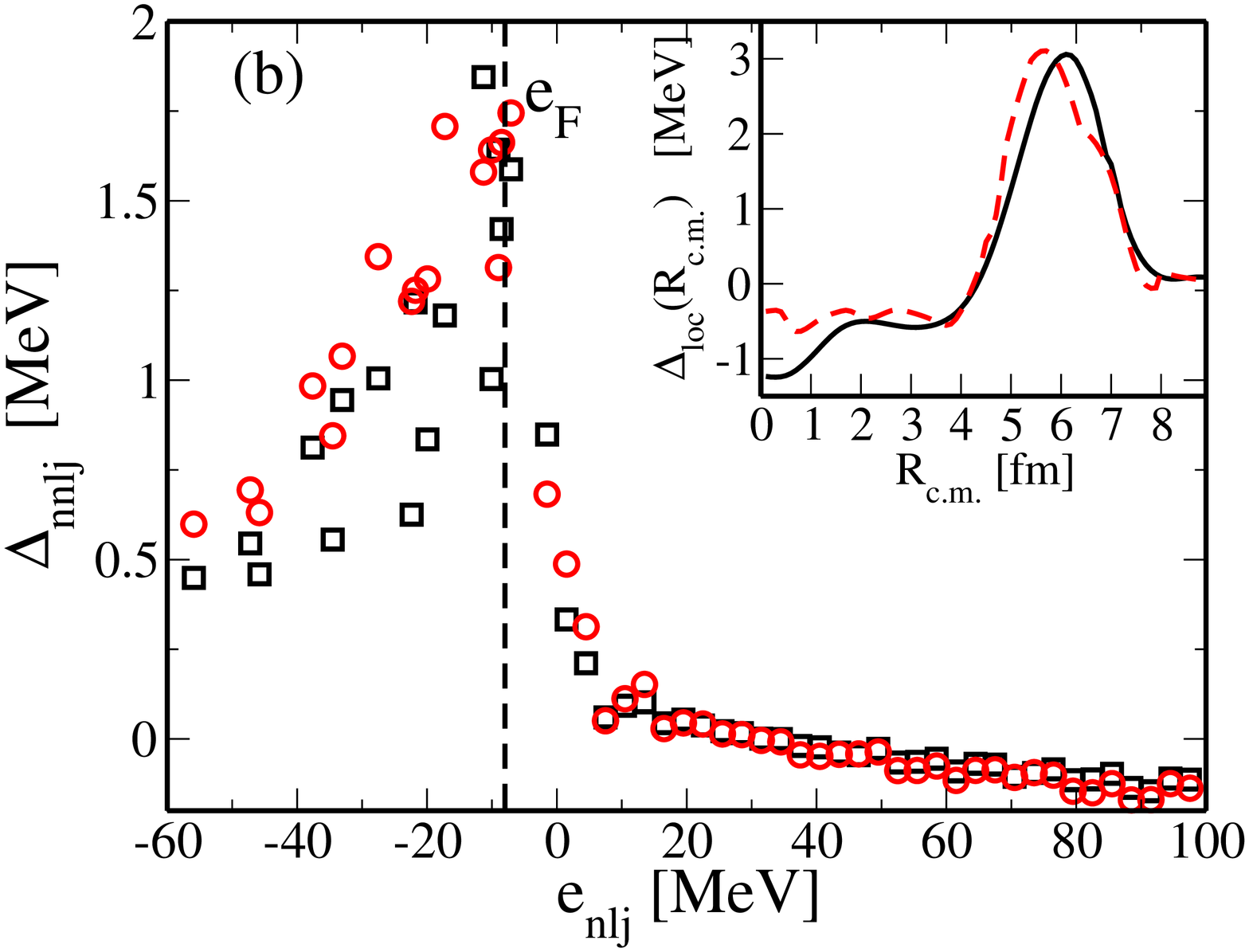}
\end{center}
\caption{\footnotesize{
The diagonal matrix elements of the pairing gap associated with the induced 
interaction $v_{ind}$ (triangles, cf. Fig.~\ref{Delta:matrix:el}) are compared with those associated with the 
Gaussian parametrization (circles, cf. Eq. \ref{Gaussian})). The spatial dependence of the 
semiclassical pairing gap associated with the induced interaction (solid line)
and with the Gaussian interaction (dashed line) are shown in the insert.
(b) The same, but for the Argonne+induced interaction $v_{Arg+ind}$, shown
by squares.}}
\label{fig:para2}
\end{figure*}

\subsection{Finite range  parametrization }

Within the zero-range parametrization discussed above, one can only try to fit
the total bare+induced interaction, and since the  resulting 
pairing interaction is a monotonic function of $R_{c.m.}$,
one cannot describe specific enhancements of the 
interaction  localized on the 
nuclear surface or within the nuclear volume. 
We shall now discuss an alternative parametrization of $v_{ind}$, 
based on the  dominantly surface  or volume character of the induced interaction associated respectively 
with the spin-independent or the spin-dependent  parts of the induced interaction.

\begin{figure}[htb]
\begin{center}
\includegraphics*[width=6cm,angle=-90]{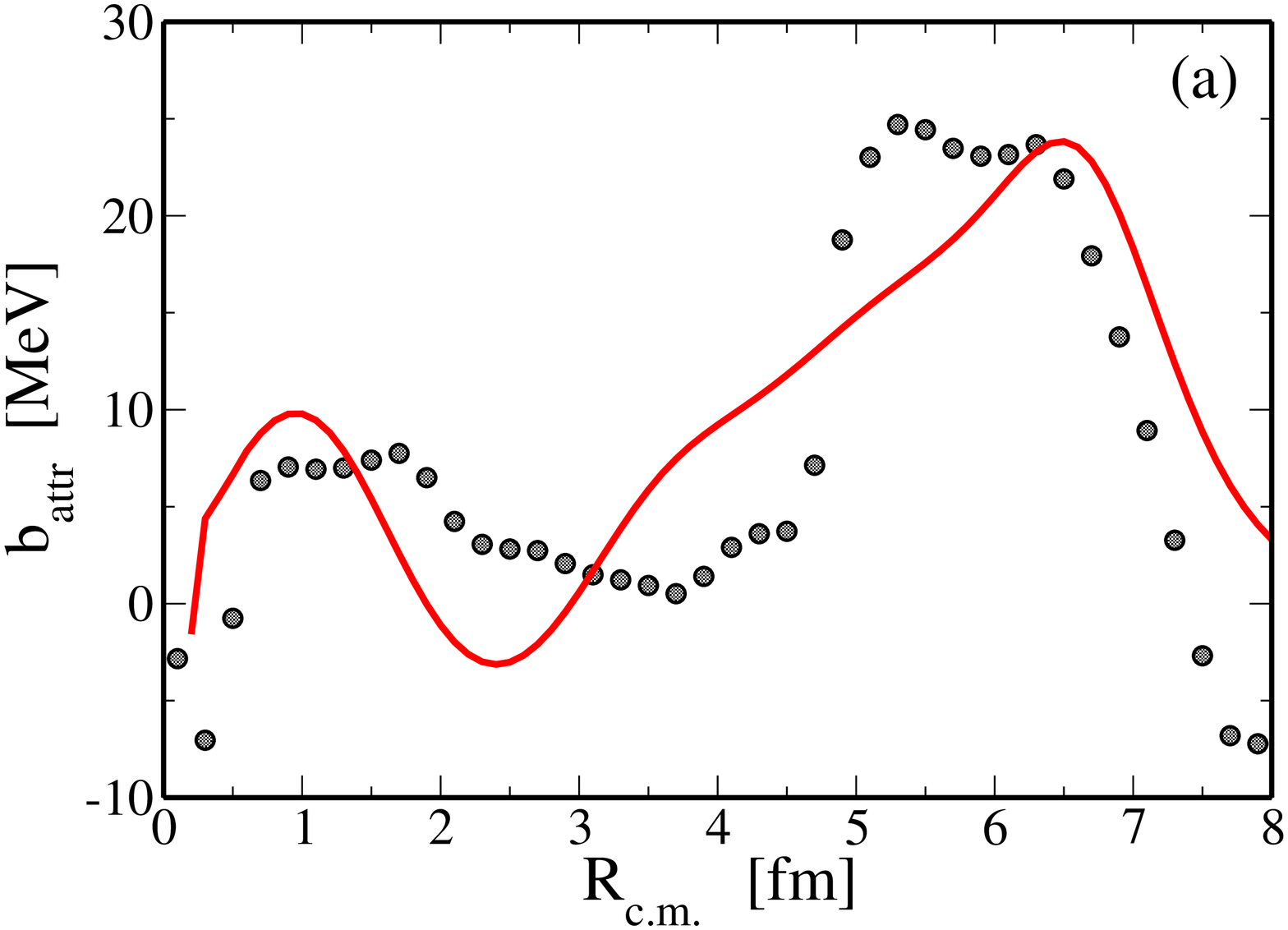}
\includegraphics*[width=6cm,angle=-90]{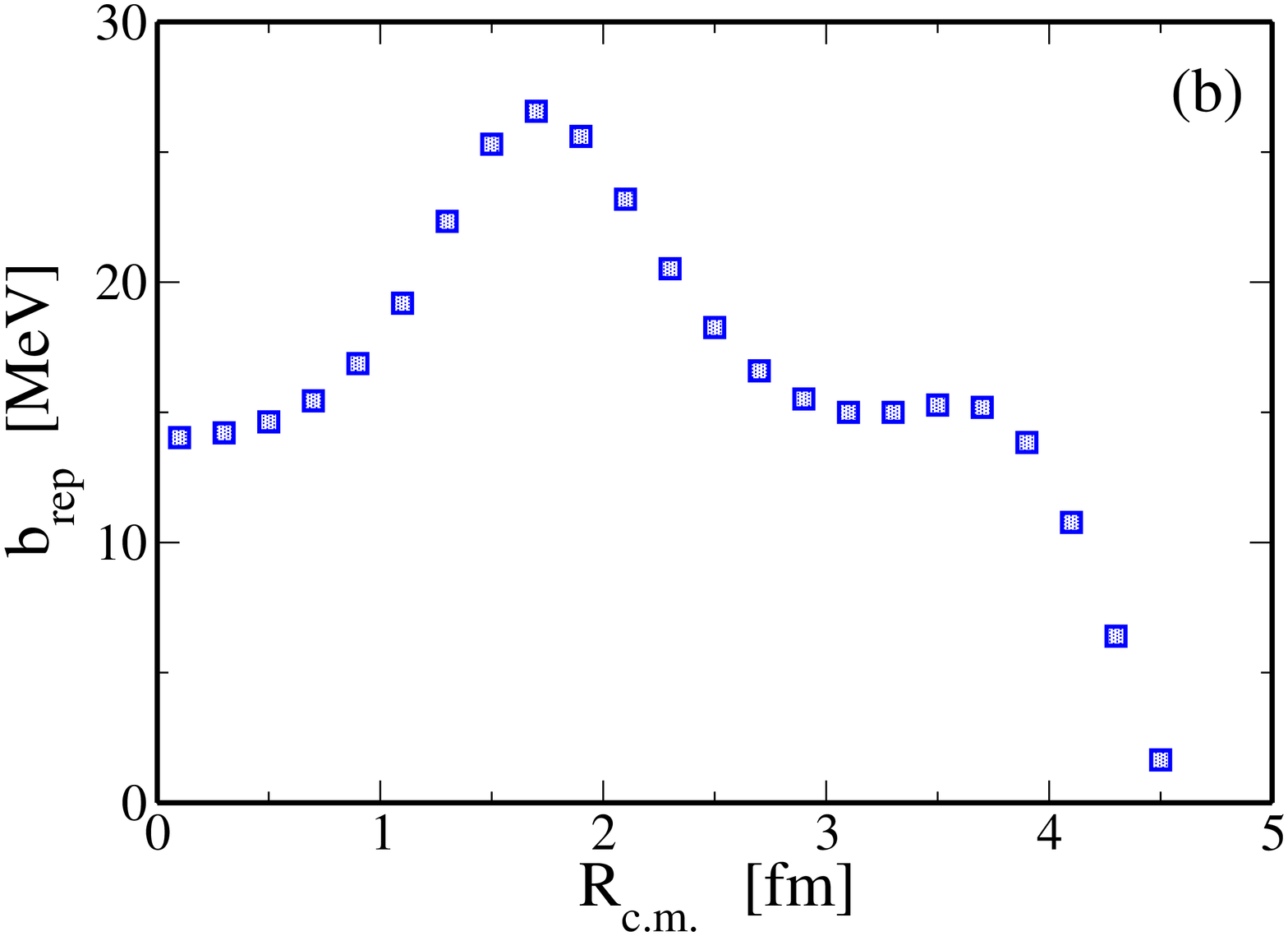}
\end{center}
\caption{\footnotesize{(a) The values of the parameter $b_{attr}$, obtained fitting  
the Gaussian interaction Eq.~(\ref{Gaussian}), are shown as a function 
of the center of mass $R_{c.m.}$ (filled dots), and are compared with 
the  function  $0.14 R_{nucl} \frac{dU(r)}{dR_{c.m.}}$ (solid line).
(b) The values of the parameter $b_{rep}$, obtained fitting  
the Gaussian interaction Eq.~(\ref{Gaussian}), are also shown as a function 
of the center of mass $R_{c.m.}$ (filled squares)}}
\label{b:fit}
\end{figure}

We shall try to determine a Gaussian function $v_{ind}^G(R_{c.m.},r_{12})$ so as
to fulfill approximately the relation 
\begin{equation}
{\Delta}(R_{c.m.},r_{12}) = - v_{ind}^G(R_{c.m.},r_{12}){\Phi}^{S=0}(R_{c.m.},r_{12}).
\label{fitg}
\end{equation}
We  consider separately the contribution from the 
spin-independent, attractive 
and spin-dependent, repulsive  parts of the interaction, writing 
$v_{ind}^G(R_{c.m.},r_{12}) = v_{attr}^G(R_{c.m.},r_{12})+ v_{rep}^G(R_{c.m.},r_{12})$.
We shall first fit the pairing gap obtained including only the spin-independent
part of $v_{ind}$ and  shown in Appendix B 
(cf.  Fig. \ref{Fig_app2}(b) and  \ref{Fig_app2}(c)), using the function
\begin{equation}\label{Gaussian}
v_{attr}^G(R_{c.m.},r_{12})= - b_{attr} \cdot {\rm exp} \left(-\left((r_{12}-c)/a_{attr}\right)^2\right) 
\end{equation}
where $a_{attr},b_{attr}$ and $c$ are parameters to be determined. 
We fix $c$ so as to constrain the  Gaussian function to be maximum
when at least one of the neutrons is on the surface of the nucleus. This implies
$c=2|R_{nucl}-R_{c.m.}|$, where $R_{nucl}=6.4$~fm is the location of the maximum  
of the first derivative of the single-particle potential. 
The parameter $a_{attr}$ turns out in all cases to be very close to  $a_{attr} \approx $ 2 fm, so in practice we have used a fixed
value $a_{attr} =$ 2 fm.
The resulting values of the parameter $b_{attr}$ obtained as a function of $R_{c.m.}$ 
are peaked on the nuclear surface  and 
are  plotted   in Fig.~\ref{b:fit}(a). They can be 
rather well reproduced by the function:
$b_{attr}(R_{c.m.}) \sim \beta_{ind} R_{nucl} \cdot\frac{dU(R_{c.m.})}{dR_{c.m.}}$, where
$\beta_{ind}=0.14$, which 
is of the order of the deformation parameter associated with the low-lying vibrational states.

The repulsive part of the induced interaction is active only in the interior of the nucleus, for
$R_{c.m.} \lesssim$  4 fm (cf. Fig \ref{Delta:Rr}(a)  and Fig. \ref{Fig_app2}(b) in Appendix B), so we multiply the Gaussian  by a Heaviside function centered at $R_0$ = 4.6 fm:
\begin{equation} v_{rep}^G(R_{c.m.},r_{12}) =
b_{rep} \cdot {\rm exp} \left(-\left(r_{12})/a_{rep}\right)^2\right)\Theta(R_{c.m.}-R_0).
\end{equation}
We then determine the parameters of the repulsive Gaussian, fitting the values 
of $a_{rep}$ and $b_{rep}$ so that the resulting 
\begin{equation}
v_{ind}^G(R_{c.m.},r_{12})=
v_{attr}^G(R_{c.m.},r_{12}) +v_{rep}^G(R_{c.m.},r_{12})
\label{vgauss}
\end{equation}
 satisfies Eq. (\ref{fitg})
for values of $r_{12}$ in the interval [0,2]~fm.
where we use in this case the gaps ${\Delta}(R_{c.m.},r_{12})$  and the Cooper pair wavefunction ${\Phi}^{S=0}(R_{c.m.},r_{12})$ obtained from the full calculation of the induced interaction cosidering both spin modes and density modes, see Fig. \ref{Phi:Rr}(a)
and \ref{Delta:Rr}(a). 
The parameter $a_{rep}$ turns out in all cases to be very close to  $a_{rep} \approx $ 3.5 fm, so in practice we have used a fixed value $a_{rep} =$ 3.5 fm. The resulting
values of $b_{attr}$ and $b_{rep}$ are shown in Fig. \ref{b:fit}(b) as a function of $R_{c.m.}$.

In Fig. \ref{fig:para2}(a) we show the diagonal matrix elements of the pairing gaps
and the semiclassical pairing gap obtained with the resulting Gaussian interaction, comparing it
with the original induced interaction. 
In Fig.~\ref{fig:para2}(b) we show instead the 
quantities obtained adding the Argonne and the Gaussian interaction in analogy to Eq.~(\ref{Int:arg+indotta}). 
One can notice that the matrix elements 
$\Delta_{nnlj}$ are better reproduced with the Gaussian interaction than with the DDDI
parametrization (cf. Fig.~\ref{fig:para1}), leading to a better agreement with
the value of $\Delta_F$ calculated with the full interaction 
(cf. Table~\ref{Tab:Gaussian}).

\begin{table}
\begin{center}
\begin{tabular}{c|c|c|c|c}
\hline
\hline
Interaction          & $\Delta_F^{full}$ & $\Delta_F^G $      & $E_{pair}^{full}$& $E_{pair}^{G}$\\
\hline
                           &                  &         & &          \\
$v_{ind}$             & 1.11 & 1.13   &   -7.41         &    -7.99       \\
$v_{Arg+ind}$         & 1.47 &  1.65  &   -15.8          &   -20.48       \\
                       & &     &                  &                  \\
\hline
\hline
\end{tabular}
\end{center}
\caption{\footnotesize{Average gaps and pairing energies (in MeV) obtained with the full
calculation and with the Gaussian parametrization $v_{ind}^G$}.}
\label{Tab:Gaussian}
\end{table}

\section{Conclusions}
The coupling of quasiparticles with collective surface vibrations gives rise to an induced
pairing interaction which  renormalizes the bare
nucleon-nucleon interaction in an important way, leading to a total pairing field which is strongly peaked at the
surface of the nucleus.  Although the pairing induced interaction is non-local
and energy dependent, it is possible to  adopt a semiclassical
approximation, which yields a local pairing field that reproduces 
to a good accuracy the features of the
full quantal solution. This local field can also be obtained 
adopting  the widely used   zero-range, density-dependent interaction, 
with an appropriate choice of the parameters, which turn out to be 
quite different from those  usually employed in more 
phenomenological approaches. 
We have also given a simple and accurate finite range parametrization of the induced interaction.

\section{Acknowledgment}
The authors wish to thank S. Baroni for providing valuable help with QRPA calculations.
F.B.  aknowledges partial support from the Spanish Education and Science Ministery projects FPA2006-13807-c02-01,
FIS2005-01105 and INFN08-33.

\clearpage

\section{Appendix A}
\vskip 0.5cm

\begin{table}
\begin{center}
\begin{tabular}{c|c|c|c|c|c|c}
\hline
\hline
Interaction                 & $\alpha$ & $\eta$ & $\Delta_F^{full}$& $\Delta_F^{\delta}$&
$E_{pair}^{full}$ & $E_{pair}^{\delta}$\\
\hline
$v_{Gogny}$                     &  0.51   & 0.63 & 1.92 & 2.05 & -20.4 & -26.6\\
$v_{Gogny}$, rescaled               & 0.38  & 0.67 & 1.46& 1.39 &-13.1 & -14.1\\
\hline
\hline
\end{tabular}
\end{center}
\caption{\footnotesize{Parameters of the DDDI, Eq. (\ref{DDDI}), 
producing pairing gaps which  fit the  
local semiclassical pairing fields obtained with the various interactions.
In the last two columns we compare the pairing gap at the Fermi energy and the pairing energy (in MeV) obtained with the full calculation, $\Delta_F^{full}$ and $E_{pair}^{full}$ with the 
values obtained using the corresponding density dependent interaction, 
$\Delta_F^{\delta}$ and  $E_{pair}^{\delta}$}.}
\label{Tab:Table_app1}
\end{table}

\begin{figure*}[!h]
\begin{center}
\includegraphics*[width=6.3cm,angle=-90]{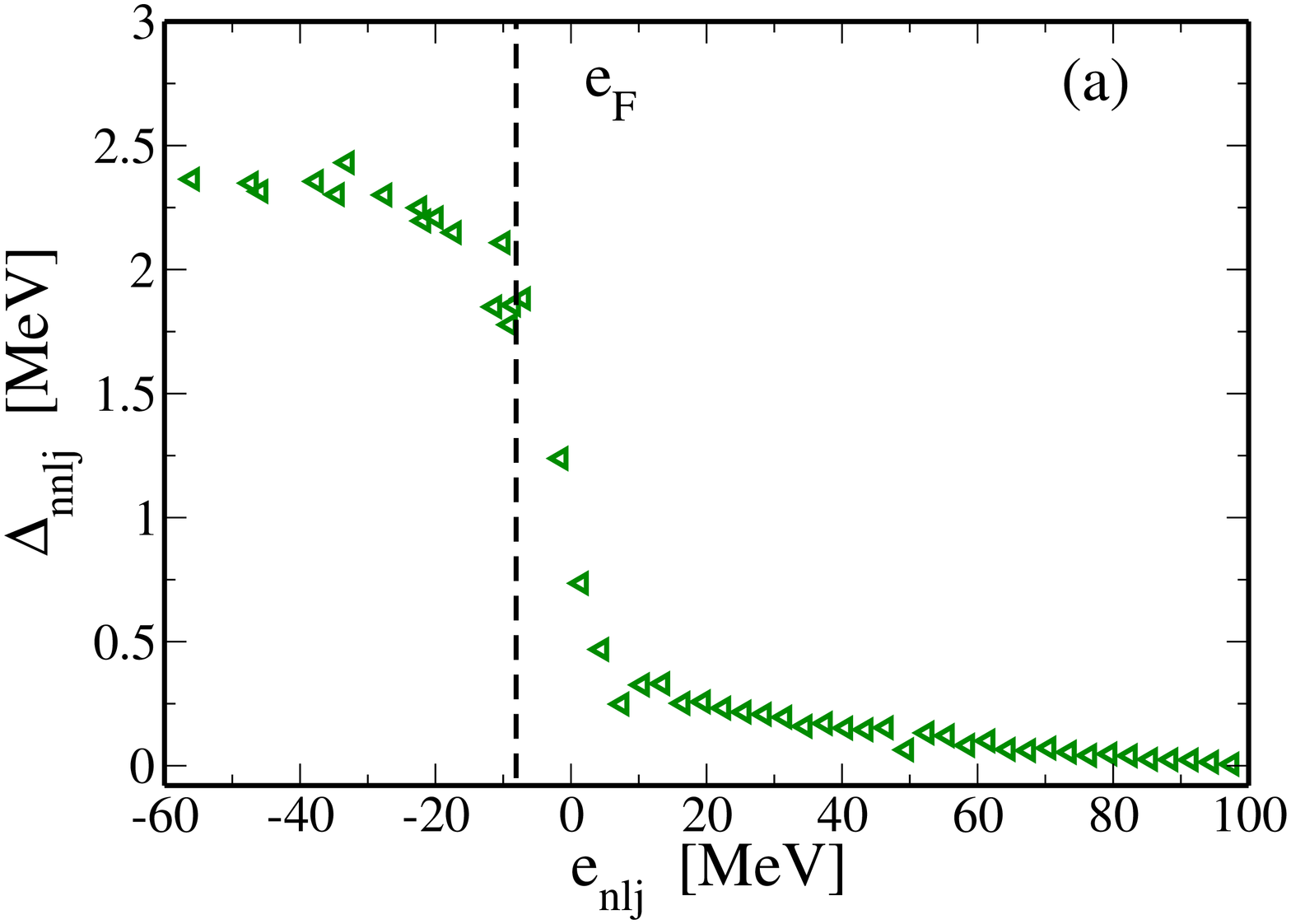}
\hspace{0.075\textwidth}
\includegraphics*[width=6.3cm,angle=-90]{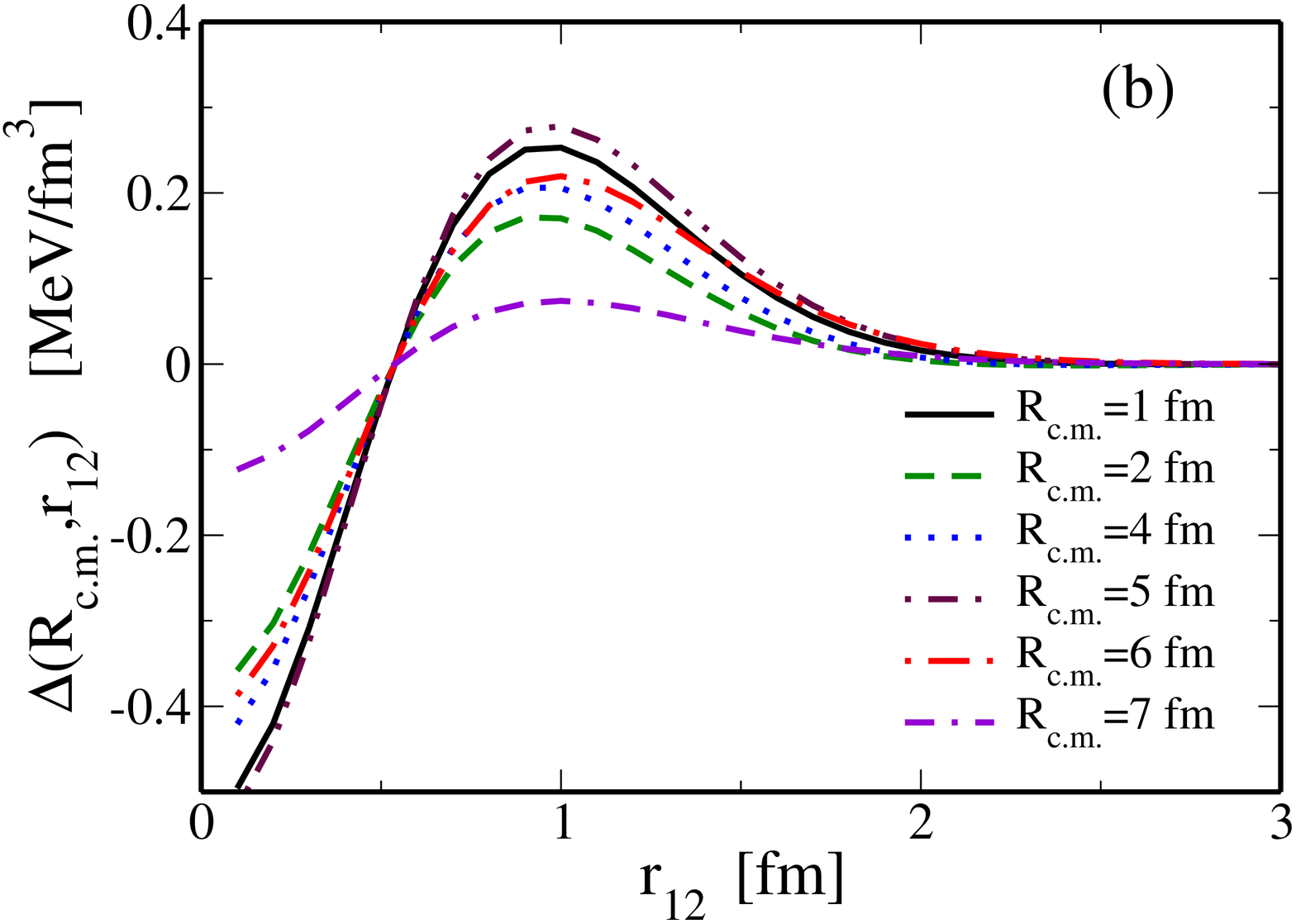}\\
\vspace{0.01\textwidth}
\includegraphics*[width=6.3cm,angle=-90]{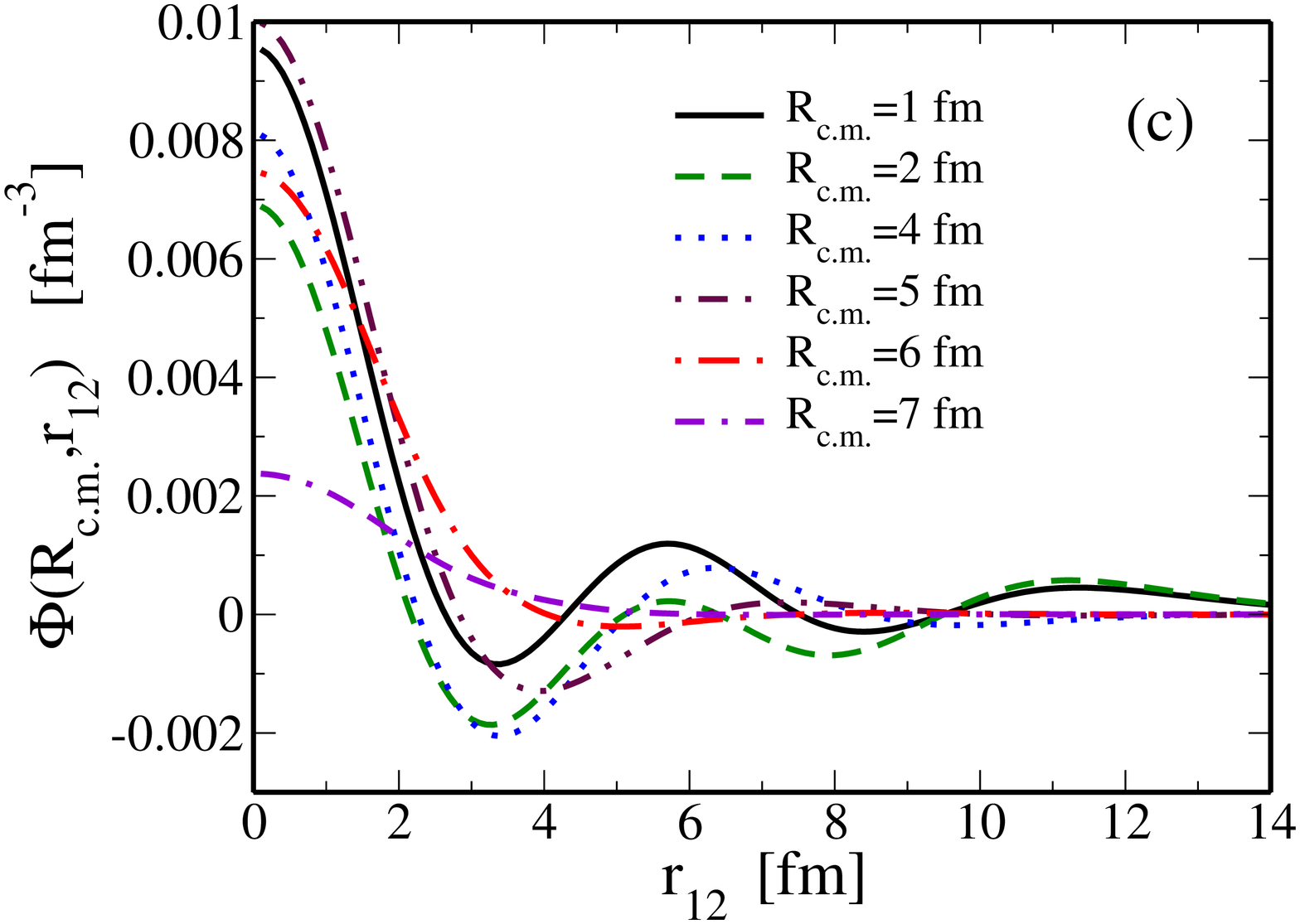}
\hspace{0.075\textwidth}
\includegraphics*[width=6.3cm,angle=-90]{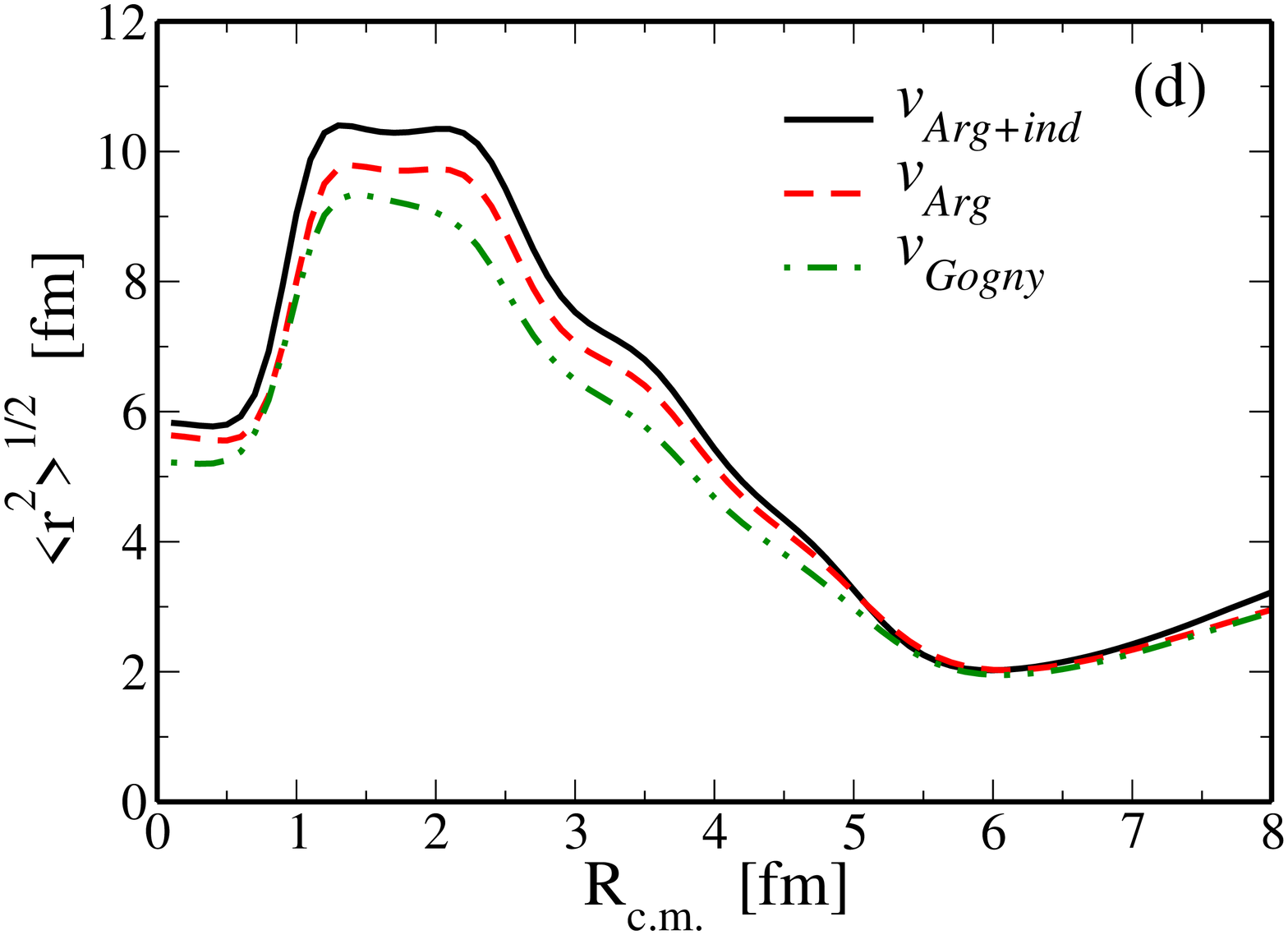}\\
\vspace{0.01\textwidth}
\includegraphics*[width=6.3cm,angle=-90]{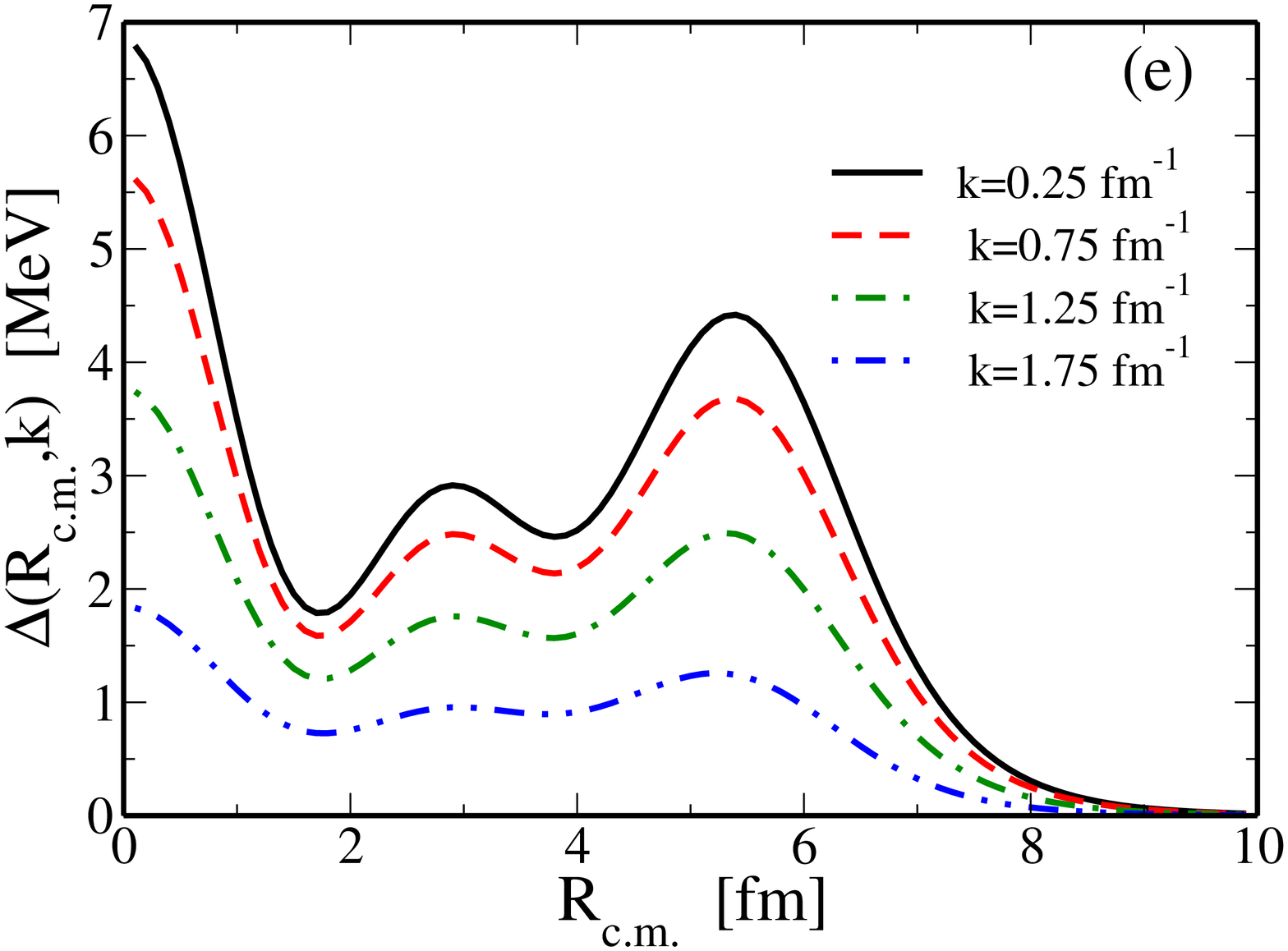}
\hspace{0.075\textwidth}
\includegraphics*[width=6.3cm,angle=-90]{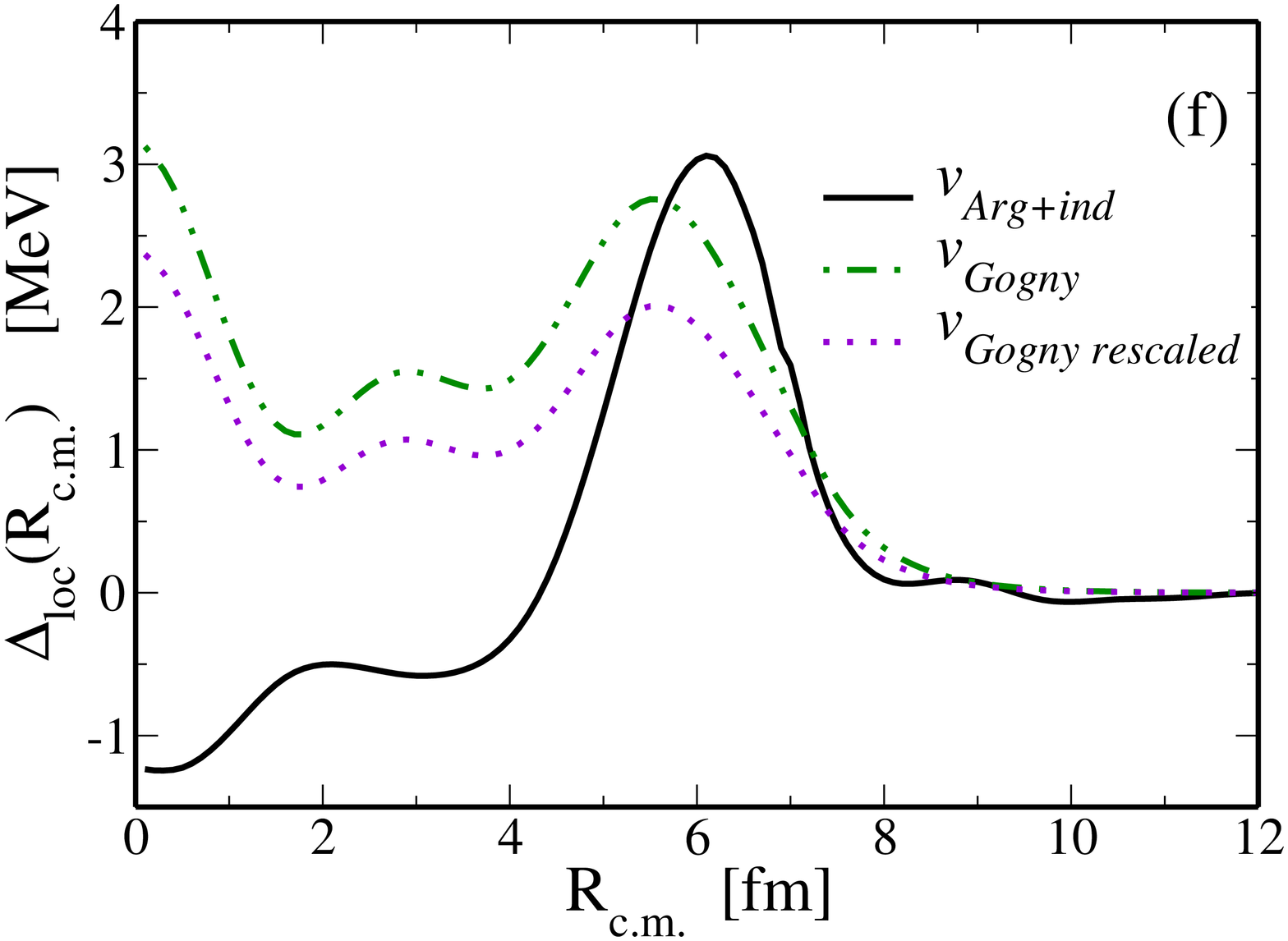} \\
\end{center}
\caption{\footnotesize{
Different pairing gaps and Cooper pair wavefunctions obtained with the Gogny interaction.
(a) Diagonal matrix elements $\Delta_{nnlj}$  as a function of the single 
particle energy $e_{nlj}$.
The vertical dashed line indicates the position of the Fermi energy.
(b) Pairing gap $\Delta(R_{c.m.},r_{12})$ in coordinate space for fixed values of $R_{c.m.}$.
(c) Abnormal density $\Phi(R_{c.m.},r_{12})$ in coordinate space for fixed values of $R_{c.m.}$.
(d) Root mean square radius of the Cooper pair as a function
of the position of the center of mass, for the Gogny interaction (dash-dotted curve),
the Argonne interaction (dashed curve) and the Argonne+induced interaction (solid curve).
(e) Pairing field (cf. Eq. (\ref{Delta:Fourier}))
as a function of the position of the center of mass
for different values of the relative 
momentum $k$.
(f) Pairing fields obtained with the semiclassical 
approximation (cf. Eq. (\ref{Delta:approx})) for the Gogny interaction (dash-dotted curve)
and for the Gogny interaction with rescaled matrix elements (dotted curve). They are compared
with the pairing field associated with the Argonne+induced interaction (solid curve), already
shown in Fig. \ref{Delta:Fermi}.}}
\label{Fig_app1}
\end{figure*}

In this Appendix, we investigate the properties of the Gogny D1S interaction. The Gogny 
interaction is an effective, finite range  interaction which reproduces rather well
the overall trends of the pairing gap along the mass table \cite{Hilaire}. 
Compared to a bare force, it has a weak repulsive core and leads to larger gaps
close to saturation density. In the following, we shall evaluate its properties in the
pairing channel, starting from the same HF field obtained with the SLy4 interaction and 
previously used. The resulting properties, however, turn out to be  similar to those
obtained in a full HFB calculation with the Gogny force. This is due to the fact that
the values of the effective mass associated with the SLy4 and Gogny interactions 
are rather similar. 
In the specific case of $^{120}$Sn, the values of its matrix elements
$\Delta_{nnlj}$, shown in Fig. \ref{Fig_app1}(a), are close to 1.8 MeV, leading to an overestimate
of the experimental gap \cite{footnote_gogny}. The pairing gap $\Delta(R_{c.m.},r_{12})$ and
the Cooper pair wavefunction $\Phi(R_{c.m.},r_{12})$ are shown in Fig. \ref{Fig_app1}(b) and \ref{Fig_app1}(c), while the
root mean square radius of the Cooper pair  is shown in Fig. \ref{Fig_app1}(d).
In Fig. \ref{Fig_app1}(e) we show the Fourier transform of the pairing field. Finally in 
Fig. \ref{Fig_app1}(f) we show the semiclassical pairing gap $\Delta_{loc}(R_{c.m.})$.  
The volume part of the interaction is considerably more pronounced compared 
to the Argonne and to the Argonne+induced interaction.

In order to compare this semiclassical gap with the analogous quantities obtained 
for the Argonne+induced interaction presented above in the main text,
we also show the semiclassical field obtained after rescaling the matrix elements
of the Gogny interaction by a factor 0.9, so as to obtain a value of the pairing gap of about 1.4 MeV at
the Fermi energy. We also show in Tab.~\ref{Tab:Table_app1} the parameters of the zero-range,
density-dependent interaction obtained fitting either the Gogny or the rescaled Gogny interaction. 

\clearpage

\section{Appendix B}\label{appendice:density}
\vskip 0.5cm 

\begin{figure*}[!h]
\begin{center}
\includegraphics*[width=6.3cm,angle=-90]{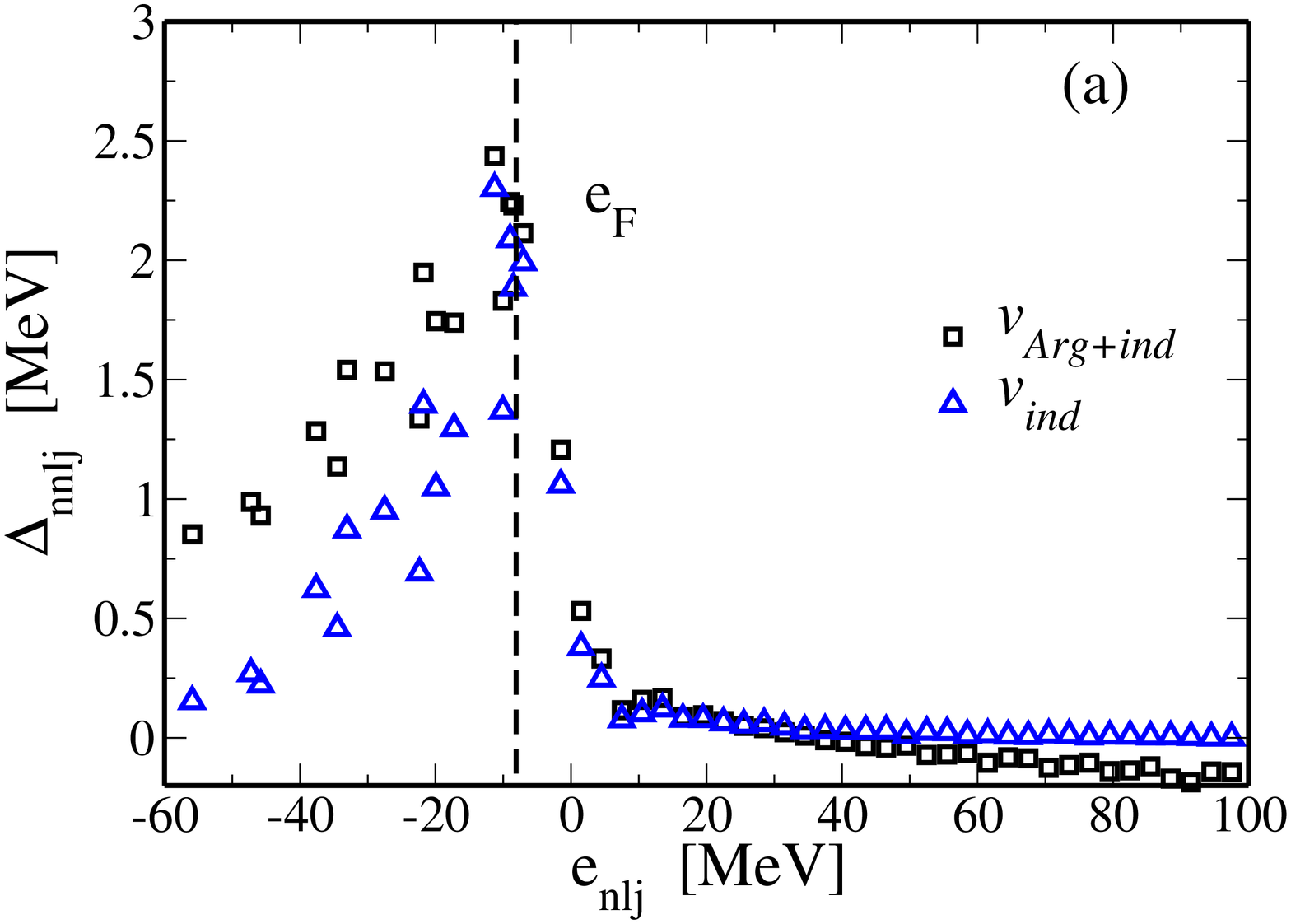}
\hspace{0.075\textwidth}
\includegraphics*[width=6.3cm,angle=-90]{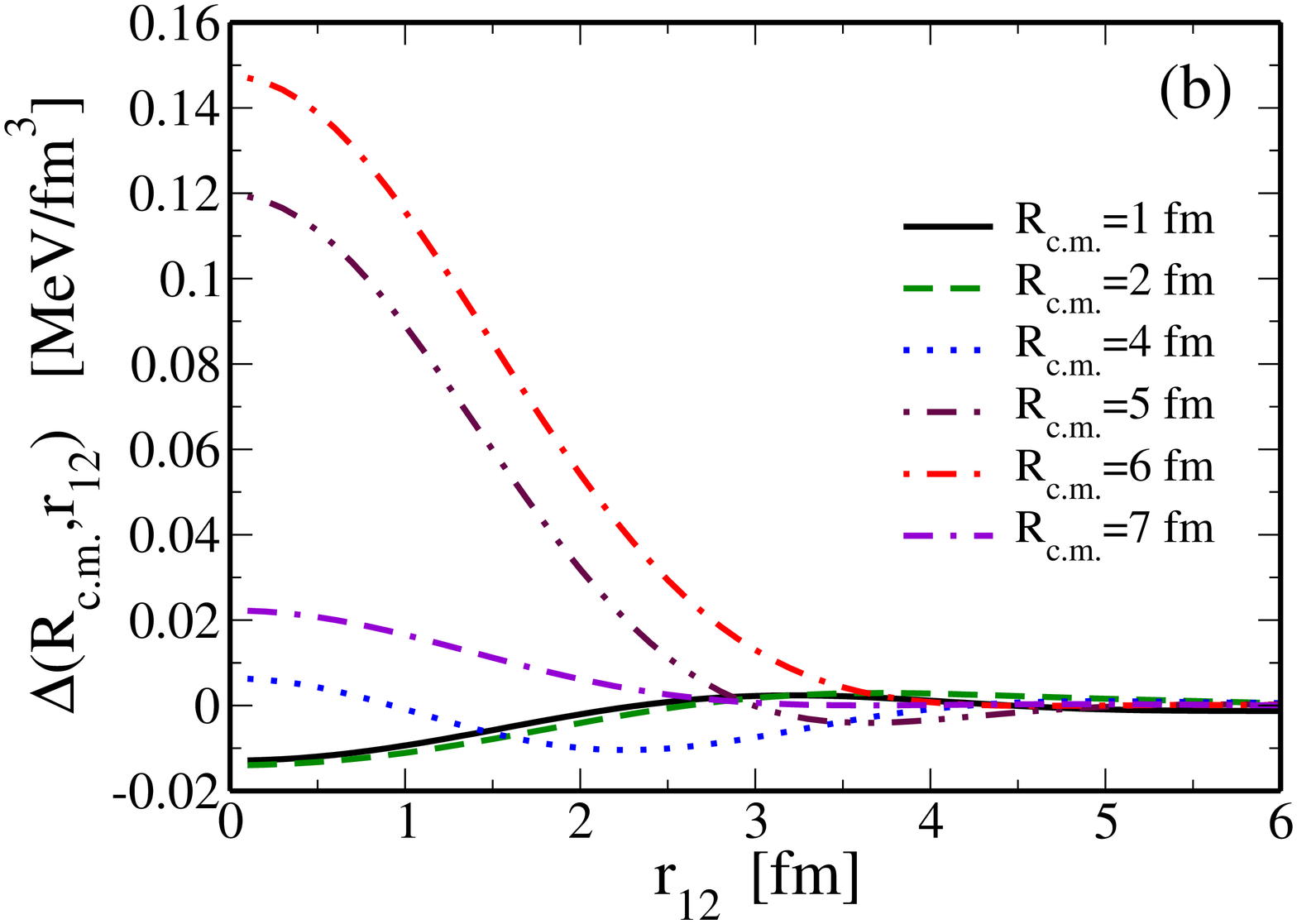}\\
\vspace{0.01\textwidth}
\includegraphics*[width=6.3cm,angle=-90]{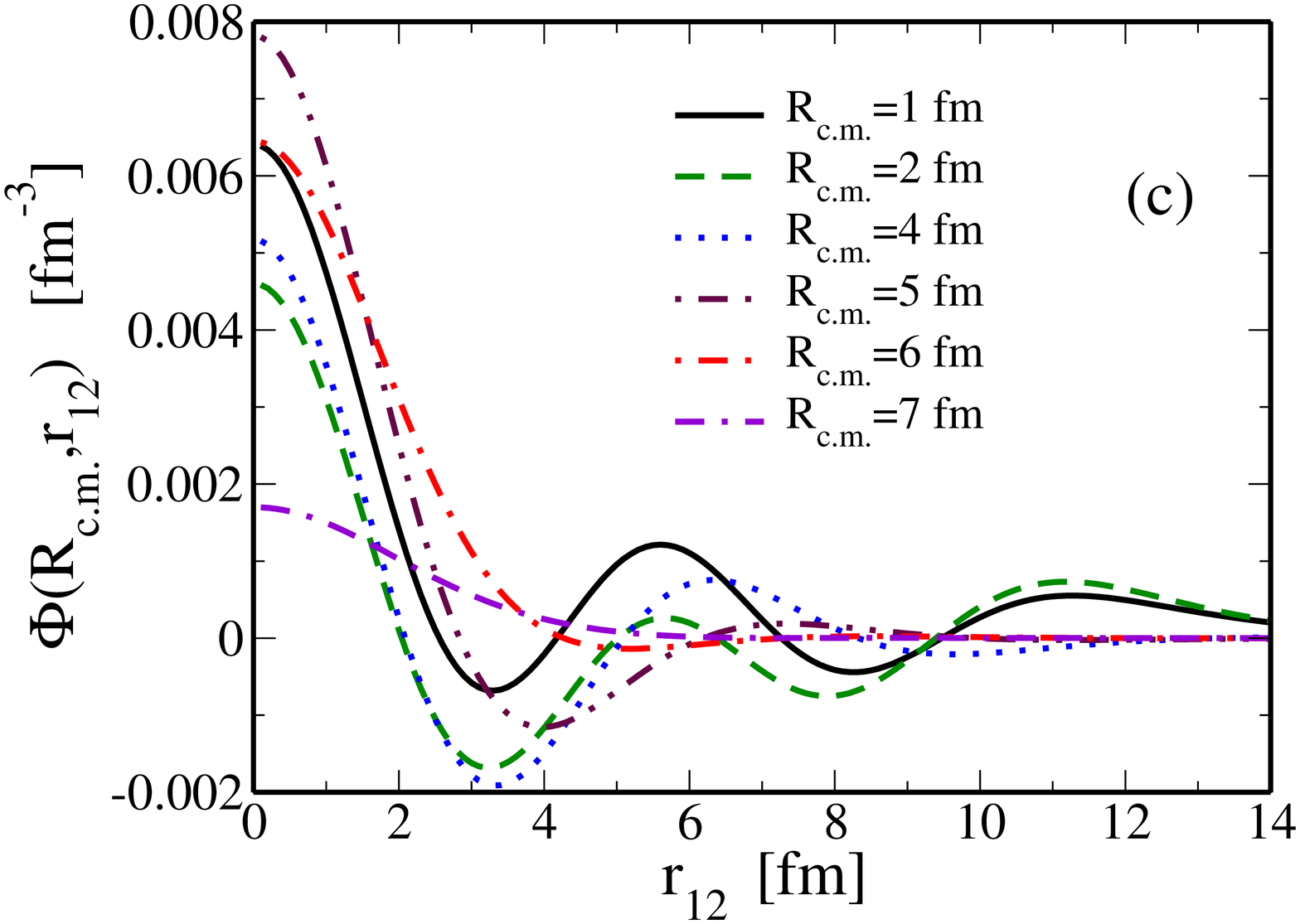}
\hspace{0.075\textwidth}
\includegraphics*[width=6.3cm,angle=-90]{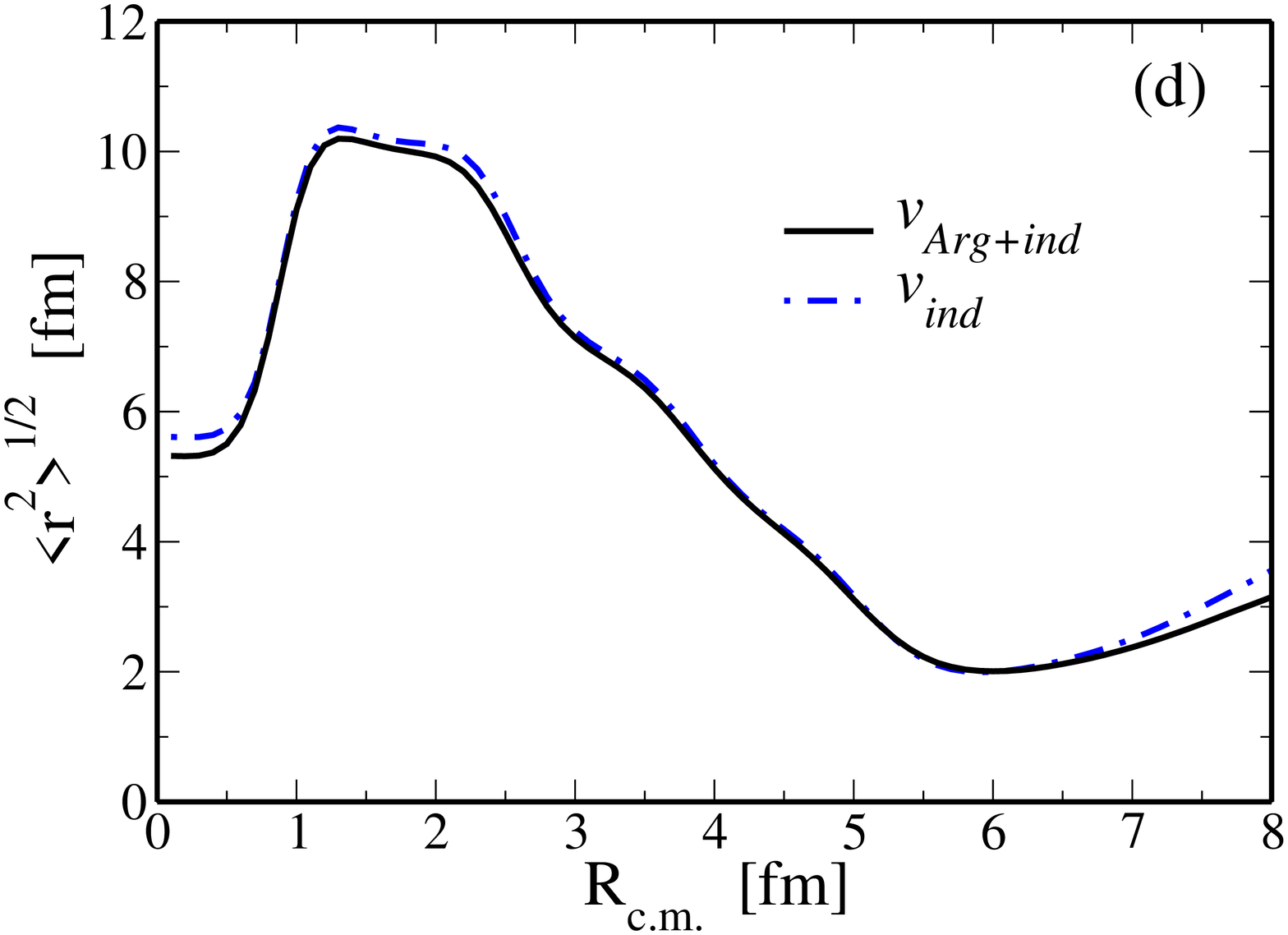}\\
\vspace{0.01\textwidth}
\includegraphics*[width=6.3cm,angle=-90]{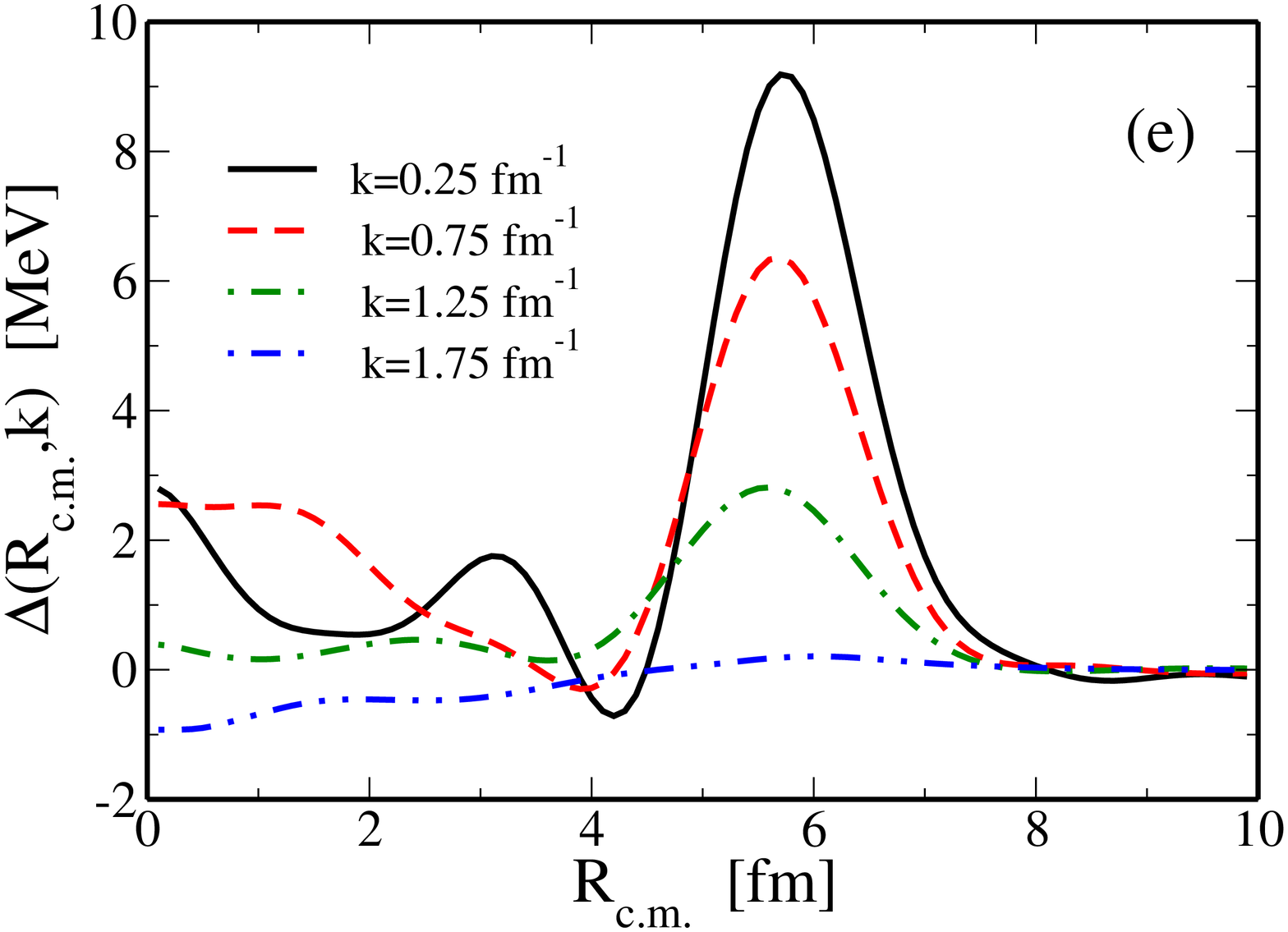}
\hspace{0.075\textwidth}
\includegraphics*[width=6.3cm,angle=-90]{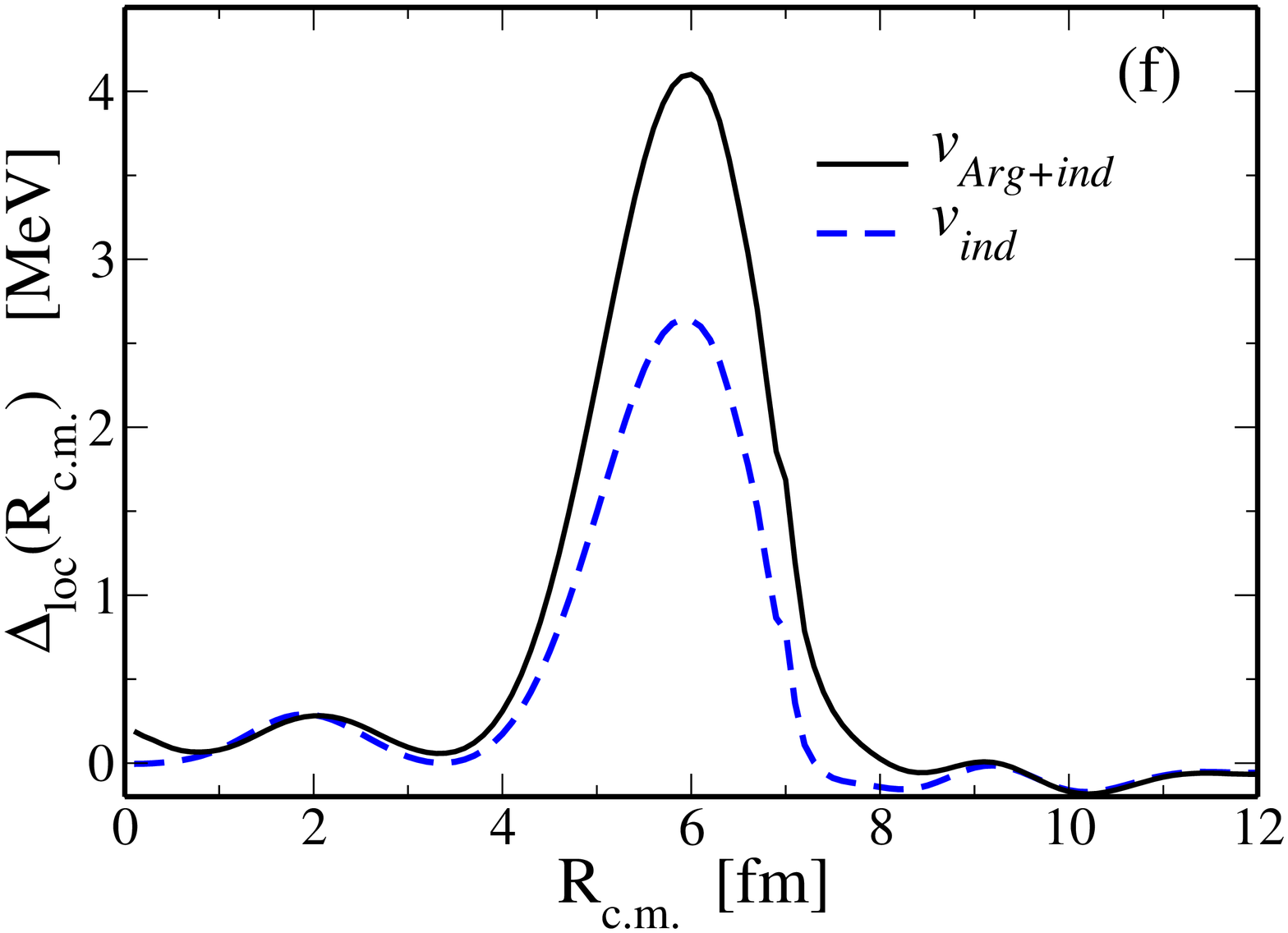} \\
\end{center}
\caption{\footnotesize{
Different pairing gaps and Cooper pair wavefunctions obtained including only 
the spin-independent part of the induced interaction.
(a) Diagonal matrix elements $\Delta_{nnlj}$  as a function of the single 
particle energy $e_{nlj}$. 
The vertical dashed line indicates the position of the Fermi energy.
(b) Pairing gap $\Delta(R_{c.m.},r_{12})$ in coordinate space for fixed values of $R_{c.m.}$.
(c) Abnormal density $\Phi(R_{c.m.},r_{12})$ in coordinate space for fixed values of $R_{c.m.}$.
(d) Root mean square radius of the Cooper pair as a function
of the position of the center of mass, for the induced interaction (dash-dotted curve),
and the Argonne+induced interaction (solid curve).
(e) Pairing field (cf. Eq. (\ref{Delta:Fourier}))
as a function of the position of the center of mass
for different values of the relative 
momentum $k$.
(f) Pairing fields obtained with the semiclassical 
approximation (cf. Eq. (\ref{Delta:approx})) for the induced interaction (dashed curve)
and for the Argonne+induced interaction (solid curve).}}
\label{Fig_app2}
\end{figure*}

In this Appendix, we show the results obtained neglecting the 
spin-dependent part  of the induced interaction, that is, setting the Landau
parameters $G_0,G'_0$ in Eq. (\ref{eqn:4}) equal to zero. In this way 
one  excludes the coupling with
non-natural modes, and produces a more attractive induced interaction. This can 
be seen for example comparing the matrix elements of the  pairing gap 
$\Delta_{nnlj}$ reported 
in Fig. \ref{Fig_app2}(a), or the pairing gap in coordinate space 
reported in Fig. \ref{Fig_app2}(b),
 with the corresponding results obtained with the full 
$v_{ind}$ (cf. Fig. \ref{Delta:matrix:el} and  Fig. \ref{Delta:Rr}). The local 
pairing gap reaches a value of 4 MeV on the nuclear surface, to be compared
with the value of  3 MeV with the full interaction (compare Fig. \ref{Fig_app2}(f)
and Fig. \ref{Delta:Fermi}). The Cooper pair wavefunction is much less  sensitive to the 
features of the interaction, as we already noticed in the main text (compare
Figs. \ref{Fig_app2}(c)-(d)
with  Figs. \ref{Phi:Rr} and \ref{Fig:phir12}).  
We show in Table \ref{Tab:Table_app2} the parameters of the zero-range density-dependent 
interaction obtained fitting the local pairing gap. 

\begin{table}
\begin{center}
\begin{tabular}{c|c|c|c|c|c|c}
\hline
\hline
Interaction                 & $\alpha$ & $\eta$ & $\Delta_F^{full}$& $\Delta_F^{\delta}$&
$E_{pair}^{full}$ & $E_{pair}^{\delta}$\\
\hline
                  & &           &       &      & & \\
$v_{Arg+ind}$             & 1.79  & 1.0 & 2.12& 2.17 & -26.6 &-31.4 \\
                            &       &      & & & & \\
\hline
\hline
\end{tabular}
\end{center}
\caption{\footnotesize{Parameters of the DDDI (cf. Eq. (\ref{DDDI})), producing pairing gaps which  fit the  
local semiclassical pairing fields obtained with the spin-independent part of the induced interaction  and with the Argonne plus induced 
interaction. 
In the last two columns we compare the pairing gap at the Fermi energy and the pairing energy (in MeV) obtained with the full calculation, $\Delta_F^{full}$ and $E_{pair}^{full}$ with the 
values obtained using the corresponding density dependent interaction, 
$\Delta_F^{\delta}$ and  $E_{pair}^{\delta}$}.}
\label{Tab:Table_app2}
\end{table}

\clearpage

\bibliographystyle{apsrev}

\begin{thebibliography}{99}

\bibitem{Brink&Broglia} D.M. Brink and R.A. Broglia \emph{Nuclear Superfluidity}, Cambridge University Press 2005.
\bibitem{Barranco_EPJA04} F.Barranco, R.A. Broglia, G. Col\`{o}, G. Gori, E. Vigezzi and P.F. Bortignon, EPJ {\bf A21} (2004) 57.
\bibitem{Barranco_99} F. Barranco, R.A. Broglia, G. Gori, E. Vigezzi, P.F. Bortignon,
J. Terasaki, Phys. Rev. Lett. {\bf 83} (1999) 2147 
\bibitem{Terasaki_02} J. Terasaki, F. Barranco, R.A. Broglia, E. Vigezzi, P.F. Bortignon,
Nucl. Phys. {\bf 697} (2002) 127 
\bibitem{Gori_Sn120}  G. Gori, F. Ramponi, F. Barranco, P.F. Bortignon, R. A. Broglia,
G. Col\`o,  E. Vigezzi, Phys. Rev. {\bf C72} (2005) 011302(R).
\bibitem{Barranco_EPA01} F. Barranco, P.F. Bortignon, R.A. Broglia, G. Col\`{o} and  E. Vigezzi, Eur. Phys. A
11(2001) 385.
\bibitem{PRC05_Barranco} F.Barranco, P.F. Bortignon, R.A. Broglia, G. Col\`{o}, P. Schuck, E. Vigezzi and X. Vi\~{n}as, Phys. Rev. {\bf C72} (2005) 054314.
\bibitem{PRC04_Gori} G. Gori, F. Barranco, E. Vigezzi and R.A. Broglia,  Phys. Rev. {\bf C69} (2004) 041302.
\bibitem{Mahaux} C.Mahaux, P. F. Bortignon, R.A.Broglia, C. H. Dasso, Phys. Rep. {\bf 120} (1985) 1.
\bibitem{Chabanat} E. Chabanat, P. Bonche and P. Haensel, Nucl. Phys. {\bf A 635} (1998) 231.
\bibitem{Colo} G. Col\`{o} and P.F. Bortignon, Nucl. Phys. {\bf A696} (2001), 427. Pairing correlations
have been calculated in the BCS approximation   with a simple seniority force, adjusted to the experimental
pairing gap of $^{120}$Sn.
\bibitem{Baldo90} M. Baldo, J. Cugnon, A. Lejeune and  U. Lombardo, Nucl. Phys. {\bf A515} (1990) 409.

\bibitem{Pillet} N. Pillet, N. Sandulescu and  P. Schuck, Phys. Rev. {\bf C76} (2007) 024310. 
\bibitem{Sandulescu_PRC05} N. Sandulescu, P. Schuck and X. Vi\~{n}as, Phys. Rev. {\bf C71} (2005), 054303.
\bibitem{Degennes} P.G. De Gennes, {\it Superconductivity of Metals and Alloys}(Addison-Wesley,
Readimg, MA, 1989). 
\bibitem{PRC98_Barranco} F. Barranco, R.A. Broglia, H. Esbensen and E. Vigezzi, Phys. Rev. {\bf C58} (1998) 1257.
\bibitem{Ring&Schuck} P.Ring and P. Schuck \emph{The nuclear many-body problem}, Springer 2005.
\bibitem{Vautherin_PRC5} D. Vautherin and D.M. Brink, Phys. Rev. {\bf C5} (1972) 626.
\bibitem{Hamamoto_PRC68} I. Hamamoto and B. Mottelson, Phys. Rev. {\bf C68} (2003) 034312.
\bibitem{Pizzochero} P.M. Pizzochero, F. Barranco, R.A. Broglia and E. Vigezzi, Astrophys. Jou.
{\bf 569} (2002) 381.
\bibitem{Bertsch_AP91} G. F. Bertsch and H. Esbensen, Ann. Phys. (N.Y.) {\bf 209} (1991) 327.
\bibitem{Garrido_PRC99} E. Garrido, P. Sarriguren, E.M. de Guerra and P. Schuck, Phys. Rev. {\bf C60} (1999) 064312.
\bibitem{Bulgac_PRL02} A. Bulgac and Y. Yu, Phys. Rev. Lett. {\bf 88} (2002) 042504.
\bibitem{Matsuo_PRC06} M. Matsuo, Phys. Rev. {\bf C73} (2006) 044309.
\bibitem{Dobaczewski_PRC06} P.J. Borycki, J. Dobaczewski, W. Nazarewicz and M.V. Stoitsov, Phys. Rev. {\bf C73} (2006) 044319.
\bibitem{Hilaire} S. Hilaire, J.-F. Berger, M. Girod, W. Satula and  P. Schuck, Phys. Lett. {\bf 531B}
(2002) 61. 
\bibitem{Doba_inter} J. Dobaczewski, W. Nazarewicz and M.V. Stoitsov, 
in {\it Proceeding of the NATO Advanced Research Workshop on the Nuclear Many-Body
Problem}, eds. W. Nazarewicz and D. Vretenar, Kluwer (2001), p. 181; 
J. Dobaczewski, W. Nazarewicz and M.V. Stoitsov, EPJ {\bf A15} (2002) 21.
 \bibitem{footnote_gogny} We note, however, that deducing the pairing gap from the calculated binding energy with the 
three-point formula would yield a better agreement.
\end{thebibliography}

\end{document}